\documentclass[12pt]{article}
\usepackage{times,graphicx,theorem,epsfig,lscape,amsmath,dcolumn,multirow,epic,float,enumerate,comment}
\usepackage{natbib}
\usepackage{booktabs}
\usepackage{amsmath,amssymb,mathrsfs,prodint,amsfonts,bbm, dsfont,bm,mathrsfs}
\usepackage[normalem]{ulem}
\usepackage{url}
\usepackage{color}

\newcommand{\blinding}[2]{#1}   

\newcommand{\R}{{\normalfont\textsf{R }}{}}
\newcommand{\lsq}{\left[}
\newcommand{\rsq}{\right]}
\newcommand{\lbc}{\left \{ }
\newcommand{\rbc}{\right \} }
\newcommand{\lp}{\left(}
\newcommand{\rp}{\right)}
\newcommand{\cond}{{\, \vert \,}}

\newcommand{\jointa}{\bar{a}_1(t), \ldots, \bar{a}_W(t)}
\newcommand{\jointA}{\bar{A}_1(t), \ldots, \bar{A}_W(t)}
\newcommand{\ja}{\bar{a}_1(t), \bar{a}_2(t)}

\begin{document}

\begin{center}
\vspace*{-2.5cm}

{\Large Estimating the causal effects of multiple intermittent treatments with application to COVID-19}

\medskip
\blinding{
Liangyuan Hu \footnote{Liangyuan Hu is corresponding author and associate professor, Department of Biostatistics and Epidemiology, Rutgers University, Piscataway, NJ 08854, USA  (email: liangyuan.hu@rutgers.edu); Jiayi Ji is Biostatistician, Department of Biostatistics and Epidemiology, Rutgers University, Piscataway, NJ 08854, USA; Himanshu Joshi is assistant professor, Institute for Health Care Delivery Science, Icahn School of Medicine at Mount Sinai, New York, New York 10029, USA; Erick R. Scott is a scientist at Kaiser Permanente Hospital Foundation, Oakland, CA 94611, USA; Fan Li is assistant professor, Department of Biostatitics, Yale University, New Haven, Connecticut 06510, USA. This research was supported in part by award ME-2021C2-23685 from the Patient-Centered Outcomes Research Institute and by grants R21 CA245855-01 and 1R01HL159077-01A1 from the National Institute of Health.} \quad \quad Jiayi Ji \quad \quad Himanshu Joshi \quad \quad Erick R. Scott \quad \quad Fan Li

}{}

\end{center}

\date{}

{\centerline{ABSTRACT}
\noindent To draw real-world evidence about the comparative effectiveness of multiple time-varying treatments on patient survival, we develop a joint marginal structural survival model and a novel weighting strategy to account for time-varying confounding and censoring. Our methods formulate complex longitudinal treatments with multiple start/stop switches as the recurrent events with discontinuous intervals of treatment eligibility. We derive the weights in continuous time to handle a complex longitudinal dataset without the need to discretize or artificially align the measurement times. We further use machine learning models designed for censored survival data with time-varying covariates and the kernel function estimator of the baseline intensity to efficiently estimate the continuous-time weights. Our simulations demonstrate that the proposed methods provide better bias reduction and nominal coverage probability when analyzing observational longitudinal survival data with irregularly spaced time intervals, compared to conventional methods that require aligned measurement time points. We apply the proposed methods to a large-scale COVID-19 dataset to estimate the causal effects of several COVID-19 treatments on the composite of in-hospital mortality and ICU admission. 

\vspace*{0.1cm}

}

\clearpage

\section{Introduction}\label{sec:intro}
The COVID-19 pandemic has been a rapidly evolving crisis challenging global health and economies. Public health experts believe that this pandemic has no true precedent in modern times \citep{oh2020covid}. While multiple COVID-19 vaccines have been developed across the globe, consensus on optimal hospital management of COVID-19 has not been achieved \citep{yousefi2020global}. In general, antiviral medications have been utilized to reduce direct pathogenic effects of viral invasion and replication, whereas immunomodulatory agents have been utilized to reduce pathogenic inflammation that has a negative impact on end-organ function.  Emerging evidence from randomized controlled trials indicates that treatment modality effectiveness varies by infection severity and patient functional status. For example, antivirals offer enhanced protection from mortality when administered near symptom onset \citep{wang2020remdesivir}, whereas immunomodulatory therapeutics such as glucocorticoids \citep{recovery2021dexamethasone} or interleukin-6 receptor antagonists \citep{horby2021tocilizumab} yield enhanced patient benefits when given in the context of severe hypoxemia requiring hospitalization and/or mechanical ventilation. Several putative therapeutics for COVID-19 have also been evaluated within randomized controlled trials.

REMAP-CAP is a randomized, embedded, multi-factorial adaptive platform trial investigating the causal effects of multiple COVID-19 treatments alone and in combination. Evidence arising from the REMAP-CAP study has provided strong evidence base for the effect of anticoagulant medications in the context of infection severity, lack of efficacy of several existing therapeutic agents (e.g. hydroxychloroquine, azithromycin, lopinavir/ritonavir, Anakinra, and therapeutic anticoagulation in critically ill patients), and long-term survival benefits from the use of antiplatelets and interleukin-6 receptor antagonists \citep{florescu2023long}.  REMAP-CAP also generated insufficient evidence of therapeutic benefit from hydrocortisone, and stronger evidence for benefits of therapeutic heparin in moderate severity infection. Further high quality evidence arising from the RECOVERY trial \citep{recovery2021tocilizumab} has demonstrated the therapeutic futility of aspirin, lopinavir/ritonavir, colchicine, and convalescent plasma. In contrast, dexamethasone, baricitinib, monoclonal antibodies, and interleukin-6 receptor antagonists, when used in conjunction with corticosteroids, have been shown to reduce the risk of mortality and/or morbidity associated with COVID-19 \citep{who2022remdesivir}. The IDSA Guidelines on the Treatment and Management of Patients with COVID-19 \citep{infectious2022idsa} provide further details of evidence-based indications and contraindications for the aforementioned therapeutics.

Although randomized controlled trials (RCTs) are considered the gold standard for establishing causal evidence of therapeutic efficacy, they are enormously expensive and time consuming,  especially in a time of crisis.  Stringent inclusion and exclusion criteria also limit the generalizability of RCTs to frailer populations  at higher risk for severe morbidity and mortality. To overcome these challenges, we study the causal effects of COVID-19 treatments on patient survival by leveraging the continuously growing observational data collected at the Mount Sinai Health System---New York City’s largest academic medical system. Treatment modalities offered at the Mount Sinai Health System during the study period aligned with ``IDSA Guidelines on the Treatment and Management of Patients with COVID-19" \citep{infectious2022idsa} and were updated as revisions were released.
We focus on four commonly used medication classes that are of most clinical interest: (i) remdesivir; (ii) dexamethasone; (iii) anti-inflammatory medications other than corticosteroids; and (iv) corticosteroids other than dexamethasone. 

\begin{figure}[H]
    \centering
    \includegraphics[width=1\textwidth]{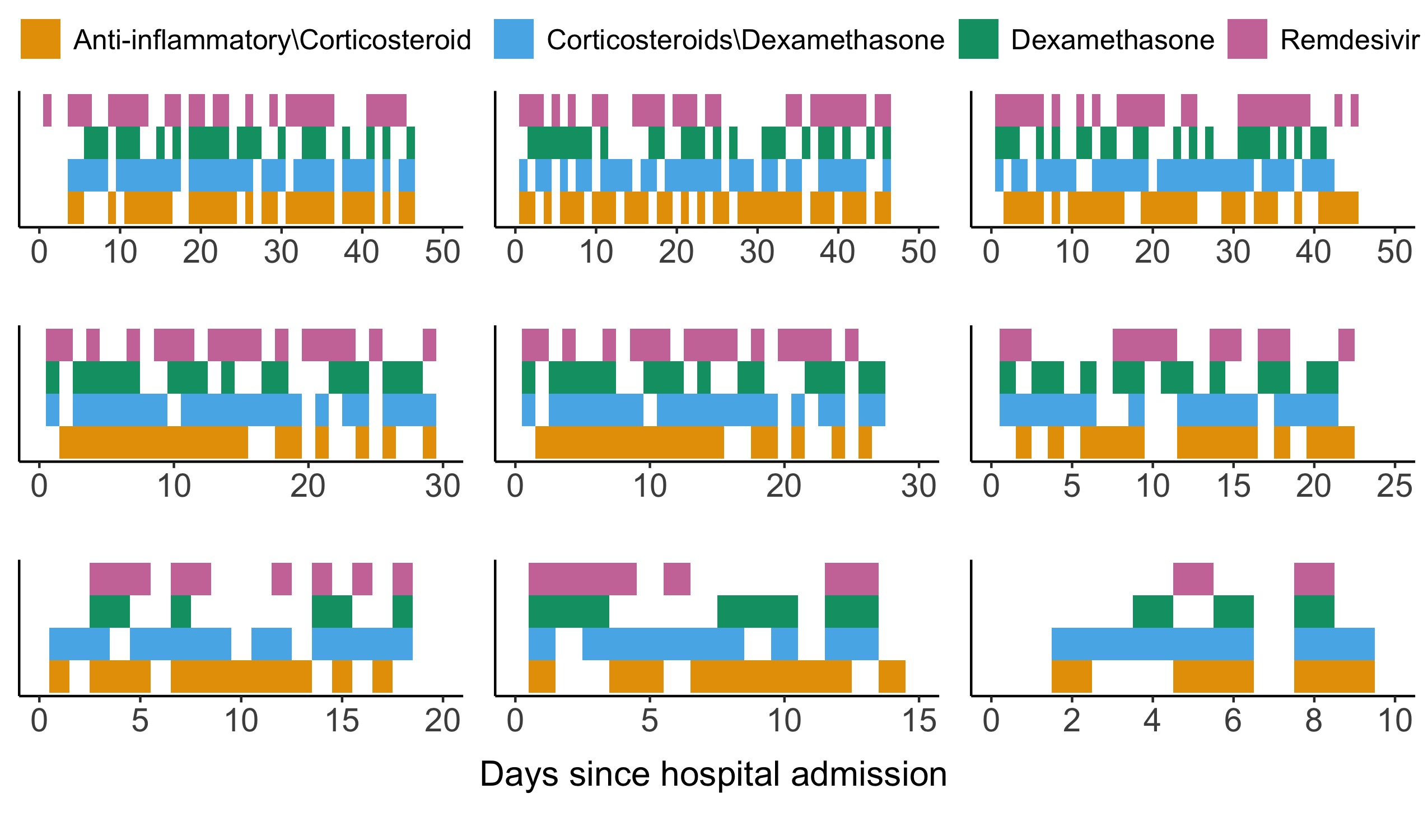}
    \caption A schematic of treatment processes for nine randomly selected patients visualized by heat maps. Colors indicate \emph{on} treatment. Lack of color corresponds to \emph{off} treatment.
    \label{fig:treat_pattern}
\end{figure}

The complex nature of COVID-19 treatments, owing to differential physician preferences and variability of treatment choices attributable to evolving clinical guidelines, poses three major challenges for statistical analysis of nonexperimental longitudinal data with censored survival outcomes. First, treatment is not randomly allocated and the treatment status over time may depend upon the evolving patient- and disease-specific covariates. Second, the measurement time points during the follow-up are irregularly spaced.  Third, there is more than one treatment under consideration.  Patients can be simultaneously prescribed to various treatment combinations, or can be switched to a different treatment.  Figure~\ref{fig:treat_pattern} illustrates the observed treatment trajectories for nine randomly selected patients during their hospital stays. Thus, the primary issue of nonrandom treatment allocation, when combined with irregularly spaced measurement time points and multiple time-varying treatments, leads to unique analytical challenges that necessitate sophisticated longitudinal causal inference approaches.

While previous work has shown that a continuous-time marginal structural model is effective in addressing time-varying confounding and provides consistent causal effect estimators \citep{johnson2005semiparametric,saarela2016flexible,hu2018modeling,hu2019causal,ryalen2020causal, johnson2022treatment}, the development has been restricted to a single longitudinal treatment and therefore may not be directly applicable. 
We consider a joint marginal structural model to accommodate multiple longitudinal treatments in continuous time. The joint effects of multiple exposures on an outcome are of strong interest in epidemiologic research \citep{rothman2008modern,howe2012estimating}. \cite{hernan2001marginal} and \cite{howe2012estimating} presented methods in the discrete time settings for estimating the joint effects of multiple time-varying exposures in HIV research. In COVID related research, there is a great need for estimating the joint or marginal effects of multiple COVID specific medications. There remains ambiguity as to which medication works best for COVID infected patients, particularly among those admitted into the hospital in earlier phases of the pandemic. In the absence of treatment guidelines, numerous medications and combinations of the medications had been used to treat COVID-19. As a result, a large repertoire of real-world clinical encounter data was generated.  A comprehensive examination of the joint or marginal effects of multiple treatments can be conducted by leveraging these datasets with a high level of granularity. This is made possible by virtue of a continuous-time \emph{joint} marginal structural model and the estimation methods we develop for such a model.  

Specifically in this article, we consider a joint marginal structural model to estimate the causal effects of multiple longitudinal treatments in continuous time. To estimate causal parameters in the joint marginal structural model, we derive a novel set of continuous-time stabilized inverse probability weights by casting each treatment process as a counting process for recurrent events, allowing for discontinuous intervals of eligibility. In addition, we propose to use machine learning and smoothing techniques designed for censored survival data to estimate such complex weights. Through simulations, we demonstrate that our approach provides valid causal effect estimates and can considerably alleviate the consequence of unstable inverse probability weights under parametric formulations. We further undertake a detailed analysis of a large longitudinal registry data of clinical management and patient outcomes to investigate the comparative effectiveness of multiple COVID-19 treatments on patient survival.

This work makes three major contributions to the causal inference literature. First, we offer a practicable framework to jointly estimate the causal effects of multiple time-varying treatments in continuous time. While methods for estimating the effect of one treatment at a time are relatively well-developed, methods for studying multiple longitudinal treatments are relatively scarce. 
Second, we develop a novel weighting scheme to deal with time-varying confounding on the continuous-time scale without the need to discretize the time and ``bin'' the data. Third, we strategically recast the problem of ``start/stop'' treatment switches as recurrent events, making it possible to draw inferences about highly complex time-varying treatments that are sparsely observed along the time continuum. Importantly, these three methodological contributions operate in concert under a single modeling framework. Thus, we provide a valuable analysis apparatus for datasets possessing highly complex structures, which are seen increasingly often as electronic health records data become increasingly accessible.

\section{Joint Marginal Structural Survival Model}\label{sec:JMSM}
\subsection{Notation and set up}
We consider a longitudinal observational study with multiple treatments and a right-censored survival outcome. Denote $t$ as the time elapsed from study entry (time 0), which is hospital admission in our COVID data example, and denote $t^o$ as the maximum follow-up time. In our COVID data example, the administrative censoring date was February 26, 2021, which defines the maximum follow-up time.
Suppose each individual has a $p$-dimensional covariate process $\{L(t): 0\leq t \leq t^o\}$, some elements of which are time-varying; by definition, the time-fixed elements of $L(t)$ are constant over $[0,t^o]$. Let $T$ denote time to an outcome event of interest such as death, 
with $\{N^T (t): 0\leq t \leq t^o\}$ as its associated zero-one counting process. We consider $W$ different medication classes (treatments), whose separate and joint causal effects on patient survival are of interest. We use $A_w(t)$ to denote the assignment of treatment $w \in \mathcal{W} =\{1, \ldots, W\}$, which can be characterized by a counting process; we let $A_w(t) =1$ if an individual is treated with $w$ at time $t$ and $A_w(t)=0$ otherwise \citep{johnson2005semiparametric}. We refer to Figure~\ref{fig:treat_pattern} for a visualization of treatment processes, in which colors indicate $A_w(t)=1$ and lack of color corresponds to $A_w(t)=0$. Due to the intricate nature of the treatment processes, we defer an exhaustive exposition on recasting the treatment process within the recurrent event framework to Section~\ref{sec:recurrent}. 

Let $C$ denote the time to censoring due to, for example, discharge or loss to follow up.  We use the overbar notation to represent the history of a random variable, for example,  $\bar{A}_w(t) = \{A_w(s): 0 \leq s \leq t\}$ corresponds to the history of treatment $A_w$ from hospital admission up to and including time $t$ and $\bar{L}(t) = \{L(s): 0 \leq s \leq t\}$ corresponds to the covariate history up to and including time $t$. Following the convention in the longitudinal causal inference literature \citep{robins2008estimation}, we assume treatment decision is made  after observing the most recent covariate information just prior to the treatment; that is,  $A_w(t)$ occurs after $L(t^-)$ for all $w$. 

Let $T^{\jointa}$ represent the counterfactual failure time to event of interest had an individual been \emph{assigned} treatment history $\{\bar{a}_1(t), \bar{a}_2(t), \ldots \bar{a}_W(t)\}$ rather than the \emph{observed} treatment history $\{\bar{A}_1(t), \bar{A}_2(t), \ldots \bar{A}_W(t)\}$. Similarly, $T^{\jointA}$ represents the observed failure time to event for an individual given the observed treatment history. We similarly define $C^{\jointa}$ as the counterfactual censoring time under treatment history $\{\bar{a}_1(t), \bar{a}_2(t), \ldots \bar{a}_W(t)\}$. The observed data available for drawing inferences about the distribution of potential outcomes are as follows: the observed time to outcome event is $T^\ast = T\wedge C$, with the censoring indicator $\Delta^T = I(T \leq C)$. Note that both treatment processes $\{A_w(t), w=1, \ldots, W\}$ and the covariate process $\bar{L}(t)$ are defined for $t\in [0,t^o]$ but are observed only at some and potentially irregularly spaced time points for each individual. For example, individual $i$ may have covariates and treatment status observed at a set of unique, discrete time points from study entry $t=0$ to his or her last follow-up time $t_{K_i} \leq t^o$. 

\subsection{Joint marginal structural model for survival outcomes}
We consider a marginal structural model to estimate the joint causal effects of multiple time-varying treatments on patient survival. The most popular model specification is a marginal structural Cox regression model, for its flexibility in handling baseline hazard and straightforward software implementation when used in conjunction with the stabilized inverse probability weights \citep{hernan2001marginal,howe2012estimating}. When there is a strong concern that the proportional hazards assumption may not be satisfied across the marginal distribution of the counterfactual survival times, alternative strategies such as the structural additive hazards model can also be considered. Other modeling frameworks such as the structural nested models \citep{robins1992estimation} can potentially be used to investigate the causal effect of a time-varying treatment on survival outcomes.  For example, \cite{lok2008statistical} presented a conceptual framework, as well as a mathematical formalization, for the application of structural nested models in drawing causal inferences about time-varying treatments in the presence of time-dependent confounding on a continuous-time scale \citep{robins1992estimation,mark1993estimating}. \citet{lok2017mimicking} further extended the theory of continuous-time structural nested models without assuming a deterministic (or constant) treatment effect. More recently, \citet{yang2020semiparametric} developed a doubly robust estimator for the continuous-time structural failure time models. However, the capacity of these methods to accommodate multiple longitudinal treatments, each characterized by multiple ``start/stop'' switches, has yet to be elucidated.

As a starting point, we present our methodology based on the marginal structural Cox regression model to address the aforementioned multiple analytical challenges (see Figure~\ref{fig:treat_pattern}), which are increasingly common with large-scale electronic health records data. However, we acknowledge that extensions based on alternative modeling frameworks such as those introduced in \citep{young2020causal} may also be possible to to overcome these challenges. Despite our focus on continuous-time marginal structural Cox model, we hope this work will open new vistas for vibrant research on this topic. 

For notational brevity but without loss of generality, we first consider $W=2$ treatments. Expansion of the joint marginal structural model and weighting scheme for $W\geq 3$ treatments is discussed in Section~\ref{sec:general}.
Specifically, we assume $T^{\bar{a}_1(t), \bar{a}_2(t)}$ follows a marginal structural proportional hazards model of the form
\begin{eqnarray} \label{eq:JMSSM}
    \lambda^{T^{\bar{a}_1(t), \bar{a}_2(t)}} (t) 
   = \lambda_0(t) \exp \lbc \psi_1 a_1 (t) + \psi_2 a_2(t) + \psi_3 a_1(t) a_2(t) \rbc,
\end{eqnarray}
where $\lambda^{T^{\bar{a}_1(t), \bar{a}_2(t)}}(t)$ is the hazard function for $T^{\bar{a}_1(t), \bar{a}_2(t)}$ and $\lambda_0(t)$ is the unspecified baseline hazard function when treatment $A_1(t)$ and $A_2(t)$ are withheld during the study. The structural parameter $\psi_1$ encodes the \emph{instantaneous} effect of treatment $A_1(t)$ (when treatment 1 is administered at time $t$) on $T^{\bar{a}_1(t), \bar{a}_2(t)}$ in terms of log hazard ratio while $A_2(t)$ is withheld during the study. Similarly, the structural parameter $\psi_2$ corresponds to the instantaneous treatment effect for $A_2(t)$ in the absence of $A_1(t)$. The instantaneous multiplicative interaction effect of $A_1(t)$ and $A_2(t)$ is captured by the structural parameter $\psi_3$, allowing the possibility that the instantaneous effect of $A_1(t)$ depends on $A_2(t)$. In addition, model~\eqref{eq:JMSSM} can be elaborated by letting $\lambda^{T^{\bar{a}_1(t), \bar{a}_2(t)}}(t)$ depend on baseline covariates or by using a stratified version of $\lambda_0(t)$, with straightforward adaptations to the weighted estimating equations (introduced in Section \ref{sec:CT-IPW}) required to identify the structural model parameters. Model \eqref{eq:JMSSM} implicitly assumes that the instantaneous treatment effect is constant across the follow-up. This model assumption is reasonable given that the COVID-related hospitalization is generally short and medications are prescribed for days in succession. 
Finally, model \eqref{eq:JMSSM} is a continuous-time generalization of the discrete-time model considered by \cite{howe2012estimating} for estimating the joint survival effects of multiple time-varying treatments. 

Model~\eqref{eq:JMSSM} offers two features for the estimation of treatment effects. First,  the counterfactual survival function can be expressed as  $$S^{T^{\bar{a}_1(t), \bar{a}_2(t)}} (t) = \exp \lbc -\int_0^t \lambda_{T^{\bar{a}_1(t), \bar{a}_2(t)}} (s)ds \rbc.$$ 
Therefore, causal contrasts can be performed based on any relevant summary measures of the counterfactual survival curves such as median survival times or restricted mean survival times. In Section~\ref{sec:analysis}, we demonstrate the  causal comparative effectiveness of COVID-19 treatments using counterfactual survival probablity at 14 days following hospital admission and 14-day restricted mean survival time. Second, model~\eqref{eq:JMSSM} allows for estimating the causal effects of static treatment regimens of complex forms as observed in clinical settings.  For instance, an intervention could be represented as $\bar{a}_1( \text{6 days)}=\bm{1}_{6\times 1}$, denoting the prescription of treatment 1 for a duration of 6 days. A more complex example is $\{\bar{a}_1(\text{12 days}), \bar{a}_2(\text{12 days})\}=\{(\bm{1}_{6\times 1},\bm{0}_{6\times 1}),(\bm{0}_{6\times 1},\bm{1}_{6\times 1})\}$, which indicates the assignment of treatment 1 for an initial 6-day period, followed by a complete switch to treatment 2 for an additional 6 days. Finally, we note that the focus of this study is treatment assignment in a population of COVID-19 patients who were admitted to the hospital following a confirmed positive diagnosis. Given the nature of the hospital and disease setting, compliance to the assigned treatment is not a major concern within our study. In the presence of noncompliance, however, our causal effects should be interpreted in an intention-to-treat manner.

\section{Estimating Structural Parameters in Continuous Time}\label{sec:CT-IPW}
To obtain a consistent estimator for $\bm{\psi}=\{\psi_1,\psi_2,\psi_3\}$ in model~\eqref{eq:JMSSM} using longitudinal observational data with two treatments, we introduce the following causal assumptions and maintain them throughout the rest of the article:

\noindent (A1) \emph{Consistency}. The observed failure times, 
$$T = \sum_{\mathcal{A}} T^{\ja} \mathbbm{1}(\bar{A}_1(t) = \bar{a}_1(t), \bar{A}_2(t) = \bar{a}_2(t)),$$  
where $\mathcal{A} = \{\bar{a}_1(t), \bar{a}_2(t): a_1(t) \in \{0,1\}, a_2(t) \in \{0,1\}, t\in [0, t^o]$\}. Similarly for the observed censoring times, 
$$C = \sum_{\mathcal{A}} C^{\ja} \mathbbm{1}(\bar{A}_1(t) = \bar{a}_1(t), \bar{A}_2(t) = \bar{a}_2(t)).$$ 
The consistency assumption implies that the observed outcome corresponds to the counterfactual outcome under a specific joint treatment trajectory $\{\bar{a}_1(t), \bar{a}_2(t)\}$ when an individual actually follows treatment $\{\bar{a}_1(t), \bar{a}_2(t)\}$. This is an extension of the consistency assumption developed with a single time-varying treatment \citep{robins1999association,wen2021parametric} to two time-varying treatments.

\noindent (A2) \emph{Conditional Exchangeability}. Alternatively referred to as \emph{sequential randomization}, this assumption states that the initiation of treatment at time $t$ among those who are still alive and remain in the study is conditionally independent of the counterfactual survival time $T^{\ja}$ conditional on observed treatment and covariate histories \citep{robins2000marginal}. Mathematically, let 
$\bar{\mathcal{O}}(t^-)=\{\bar{L}(t^-), \bar{A}_1(t^-), \bar{A}_2(t^-)\}$ denote the observed history up to $t^-$, then $\forall~t\in [0,t^o]$
\begin{equation} \label{eq:Exchange}
\rho^{A_1,A_2}\lp t \cond \bar{\mathcal{O}}(t^-), T>t^-,C>t^-,T^{\ja}\rp = \rho^{A_1,A_2}\lp t \cond \bar{\mathcal{O}}(t^-), T>t^-,C>t^-\rp,
\end{equation}
where $\rho^{A_1,A_2}(t)$ is the joint intensity of the joint counting processes defined by $A_1(t)$ and $A_2(t)$. In a similar manner, we maintain the assumption of  conditional exchangeability for censoring.
The assumption entails that $\forall t\in [0,t^o]$, 
\begin{equation}
    \lambda^C\lp t \cond \bar{\mathcal{O}}(t^-), T>t^-,C>t^-,T^{\ja}\rp
    =\lambda^C\lp t \cond \bar{\mathcal{O}}(t^-), T>t^-,C>t^-\rp .
\end{equation}
 Our conditional exchangeability assumption is a continuous-time generalization of the usual sequential randomization assumption for the discrete-time marginal structural models \citep{robins1999association,howe2012estimating}.

\noindent (A3) \emph{Positivity}. We assume that at any
given time $t$, there is a positive probability of initiating a treatment, among those who are eligible for initiating at least one treatment, for all configurations $\bar{\mathcal{O}}(t^{-})$:
$$P \lbc \rho^{A_1,A_2} (t \cond \bar{\mathcal{O}} (t^-), T > t^-, C>t^-) > 0\rbc = 1.$$ For a pair of joint treatments $(A_1, A_2)$, at a given time $t$, individuals with treatment status $(0,0), (0,1)$ or $(1,0)$ are ``eligible for'' initiating at least one treatment. 
Individuals with treatment status $A_1(t)=1$ and $A_2(t)=1$ are by nature not eligible for initiating either $A_1$ or $A_2$ at time $t$ (see our recurrent event framework in Section~\ref{sec:recurrent}). For this reason, we only need to assume the positivity assumption when the individual is \emph{off} that specific treatment, i.e., at risk for initiating that treatment. To refrain from adding complexity to the already intricate methodology, we do not consider treatment discontinuation (from 1 to 0) as a stochastic process. This decision also is justified clinically because  COVID medication is typically prescribed for a specific duration, e.g., administer dexamethasone at a doseage of 6 mg once daily for a duration of 10 days \citep{recovery2021dexamethasone}. In Section~\ref{sec:mod-assump-est}, we examine and discuss the validity of these structural (nonparametric) assumptions, along with their implications, within the context of our COVID-19 data application. 

\subsection{Framing repeated treatment initiation as recurrent events}\label{sec:recurrent}

As depicted in Figure~\ref{fig:treat_pattern}, the observed treatment patterns for COVID-19 demonstrate a high degree of complexity, owing to considerable variations in established treatment protocols and the heterogeneity in clinician preferences during the course of the pandemic. Individuals might discontinue a particular treatment and subsequently resume it at a later point, or they could be transitioned entirely to an alternative treatment.  Patients may also receive multiple treatments simultaneously for a specific duration.  From the counting process perspective, each treatment can  be viewed as a recurrent event process, with discontinuous intervals of treatment eligibility \citep{andersen1982cox}. Specifically, reframing treatment initiation as recurrent events can effectively capture two key aspects of our observational data: (i) once a patient receives a treatment, they cannot be prescribed the same treatment again while they are still \emph{on} that treatment; and (ii) after the patient is \emph{off} the treatment, they become eligible, or \emph{at risk} for re-initiating the treatment. Note that the off-treatment period is represented by the lack-of-color period for a treatment in Figure~\ref{fig:treat_pattern}. 

To formalize the treatment initiation process, we first consider a univariate treatment process $N^{A_w}$. We assume that jumps in $A_w(t)$, i.e., $dA_w(t)$, are observed on certain subintervals of $[0,t^o]$ only. Specifically for individual $i$, we observe the stochastic process $A_{w,i}(t)$ on a set of intervals 
$$\mathcal{E}_{w,i} = \bigcup_{j=1}^{J_i} (V_{w,ij}, U_{w,ij}],$$
where $0 \leq V_{w,i1} \leq U_{w,i1} \leq \ldots \leq V_{w,iJ_i} \leq U_{w,iJ_i} \leq t_{w,iK_i}$. Implications of this representation concerning treatment initiation timing are provided in Supplementary Section 3. 

Define a censoring or filtering process by $D^{A_w}_i(t) = I(t \in \mathcal{E}_{w,i})$, and the filtered counting process by 
$N^{A_w}_{iq}(t) = \int_0^t D^{A_w}_i(u) dA_{w,iq}(u),$
where $q$ indexes the $q$th treatment initiation. We assume conditional independence among occurrences of treatment initiation given all observed history \citep{andersen1982cox}, and that the set $\mathcal{E}_{w,i}$ is defined such that $D_i^{A_w}(t)$ is predictable. 
The assumption of conditional independence is considered plausible in the context of COVID-19 treatment, as the associations between treatment initiations for each patient are more likely driven by key time-varying covariates (e.g., oxygen levels). Consequently, by conditioning on these time-varying covariates, the time intervals between treatment initiations become conditionally uncorrelated, supporting this conditional independence assumption for this particular application. The observed data with occurrences on the set $\mathcal{E}_{w,i}$ can therefore be viewed as a \emph{marked point process} generating the filtration $(\mathscr{F}_{w,t}^D)$ \citep{andersen1993statistical}. Similarly, we denote the filtration generated by the counting process $\{A_w(t):t\in[0,t^o]\}$ corresponding to  $\mathcal{E}_{w,i}\in [0,t^o]$ by $(\mathscr{F}_{w,t})$. We assume $A_{w,iq}(t)$ follows Aalen's multiplicative intensity model \citep{aalen2008survival}, 
$$\rho_{w,iq}(t,\theta) = \alpha_{iq}(t,\theta) Y_{w,iq}(t),$$ 
with respect to the filtration $(\mathscr{F}_{w,t})$. In this model, the intensity process of $A_{w,iq}(t)$ is denoted as $\rho_{w,iq}(t,\theta)$. The person-specific initiation intensity, parameterized by $\theta$, is represented by $\alpha_{iq}(t,\theta) = \alpha_0(t)\text{IR}(\theta,\bar{L}_i(t))$, where $\alpha_0(t)$ corresponds to a common baseline intensity. Furthermore, the intensity ratio function for treatment initiation, which depends on the person-specific history, is given by $\text{IR}(\theta,\bar{L}_i(t))$. The at-risk function, $Y_{w,iq}(t)$, is defined as follows: $Y_{w,iq}(t) = 1$ indicates that person $i$ is eligible for the $q$th initiation of treatment $w$ just before time $t$ within the interval $[t, t+dt)$, while $Y_{w,iq}(t) = 0$ implies ineligibility. This model assumes that the intensity of initiating treatment $w$  at time $t$ can be decomposed into a product of the person-specific intensity function and the at-risk process. Within the context of COVID-19 treatment, the intensity function represents the propensity for receiving treatment $w$, attributable to (time-fixed and time-varying) factors such as disease aggressiveness and genetic predisposition. Concurrently, the at-risk process represents patients at time $t$ who have not yet received and are eligible for treatment $w$. It follows that the filtered counting process $N_{iq}^{A_w}(t)$ follows the multiplicative intensity model
\begin{equation} \label{eq:uni_inten_mod}
    \rho^{A_w}_{iq}(t,\theta) = \alpha_{iq}(t,\theta) Y_{iq}^{A_w}(t)
\end{equation}
with respect to $(\mathscr{F}_{w,t}^D)$ \citep{andersen1993statistical}. Here, $Y_{iq}^{A_w}(t) = Y_{w,iq}(t) D_i^{A_w}(t)$. With two treatments, model~\eqref{eq:uni_inten_mod} can be directly extended for the joint treatment initiation process as 
\begin{equation} \label{eq:multv_inten_mod}
    \rho^{A_1,A_2}_{iq}(t,\theta) = \alpha_{iq}(t,\theta) Y_{iq}^{A_1,A_2}(t),
\end{equation}
where $Y_{iq}^{A_1,A_2}(t)$ 
is the at-risk process for the $q$th treatment initiation with the filtering process jointly defined by $A_1$ and $A_2$. 

\subsection{Derivation of the continuous-time weights}\label{sec:weights}
We first consider the case without right censoring. Under assumptions (A1)-(A3), a consistent estimator of $\bm{\psi}$ can be obtained by solving the weighted partial score equation \citep{johnson2005semiparametric,hu2018modeling},
\begin{equation} \label{eq:w_ee}
     \sum_{i=1}^n \int_0^\infty \Omega^{A_1,A_2}(t_{K_i})  \lbc Z(A_{1i},A_{2i},t) - \bar{Z}^{\ast }(t;\bm{\psi}) \rbc dN_i^T(t)=0, 
\end{equation}
where $\Omega^{A_1,A_2}(t_{K_i})$ is the weight that corrects for time-varying confounding for time-varying treatments $A_1$ and $A_2$, $Z(A_{1i},A_{2i},t)_{(3\times 1)} = [A_{1i}(t), A_{2i}(t), A_{1i}(t)A_{2i}(t)]^{\top}$,  and
\begin{equation}\label{eq:weightedZ}
\bar{Z}^{\ast} = \dfrac{\sum_{k \in \mathcal{R}_t^T} Z(A_{k1}, A_{k2}, t) Y^{\ast T }_{k}(t) r(A_{k1},A_{k2},t ;\bm{\psi})}{\sum_{k \in \mathcal{R}_t^T}Y^{\ast T}_{k}(t) r(A_{k1},A_{k2},t ;\bm{\psi}) }
\end{equation}
is a modified version of the weighted mean of $Z$ over observations still at risk for the outcome event at time $t$ ($\mathcal{R}_t^T$ is the risk set at time $t$). In equation \eqref{eq:weightedZ}, we define the weighted at-risk indicator for outcome $Y^{\ast T}_{i}(t) = \Omega^{A_1,A_2}(t_{K_i}) Y_i^T(t)$, where $Y_i^T(t)$ is the at-risk function for the outcome event, and $r(a_1,a_2,t) = \exp \{\psi_1 a_1(t) + \psi_2 (t) a_2(t) + \psi_3 a_1(t) a_2(t)\} $. In Supplementary Section 4, we provide a heuristic justification for the consistency of the weighted estimating equations approach based on the use of Radon-Nikodym derivative \citep{murphy2001marginal,hu2018modeling}. 

In the discrete-time setting with non-recurrent treatment initiation, the stabilized inverse probability weights (we suppress subscript $i$ for brevity) are given in the prior literature \citep{hernan2001marginal,howe2012estimating}: 
\begin{align} \label{eq:dt-ipw}
\begin{split}
    \Omega^{A_1,A_2}(t) = & \lbc \prod_{\{k: t_k \leq t\}} \dfrac{P \lp A_1(t_k)=a_1(t_k) \cond \bar{A}_1({t_{k-1}}), \bar{A}_2({t_{k-1}}) \rp }{P \lp A_1(t_k) = a_1(t_k) \cond \bar{A}_1({t_{k-1}}), \bar{A}_2({t_{k-1}}), \bar{L}(t_{k-1}), T \geq t, C \geq t  \rp} \rbc \times \\
    & \lbc \prod_{\{k: t_k \leq t\}} \dfrac{P \lp A_2(t_k) = a_2(t_k) \cond \bar{A}_1({t_{k}}), \bar{A}_2({t_{k-1}}) \rp }{P \lp A_2(t_k)=a_2(t_k) \cond \bar{A}_1({t_{k}}), \bar{A}_2({t_{k-1}}), \bar{L}(t_{k-1}), T \geq t, C \geq t  \rp}\rbc,
    \end{split}
\end{align}
where $t_k$'s are a set of ordered discrete time points common to all individuals satisfying $0=t_0 < t_1 <t_2<\ldots \leq t$. While $\Omega^{A_1,A_2}(t)$ in \eqref{eq:dt-ipw} corrects for time-varying confounding by adjusting for $\bar{L}(t)$ in the weights, it requires that the time points are well aligned across all individuals. In addition, it does not accommodate the recurrent nature of complex interventions as in our observational study.

We now generalize the discrete-time weights~\eqref{eq:dt-ipw} to the continuous‐time setting so that the methods can be applied to data with irregularly spaced time intervals without artificially aligning time points.  We first partition the time interval $[0,t]$ into a number of small time intervals, and let $dA_w(s)$ be the increment of $A_w$ over the small time interval $[s, s + ds), \forall s \in [0,t ]$. Recall that treatment initiation, or the jumps of $A_w(t)$, $dA_w(t)$, is 
observed on a number of subintervals of $[0, t^o]$ only. That is, conditional on history $\bar{L}(s)$, the occurrence of treatment initiation for an individual  in $[s, s + ds)I(s \in \mathcal{E})$ is a  Bernoulli trial with outcomes $dA_w(s) = 1$ and $dA_w(s) = 0$. Then the term, $P\lp A_w(t_k) = a_w(t_k) \cond \bullet \rp$, in equation~\eqref{eq:dt-ipw} can be represented by 
\begin{equation*}
D^{A_w}(s)\lbc P ( dA_w(s)= 1 \cond \bullet ) \rbc^{dA_w(s)}\lbc P (dA_w(s)= 0 \cond \bullet ) \rbc^{1-dA_w(s)}, 
\end{equation*}
which takes the form of the individual partial likelihood for the filtered counting process $\{D^{A_w}(s) A_w (s): 0 \leq s \leq t \}$. When the number of time intervals in $[0,t]$ increases and $ds$ approaches zero, the finite product over the number of time intervals of the individual partial likelihood will approach a product integral \citep{aalen2008survival}, that is, 
\begin{eqnarray} \label{eq:prod_int}
  &&\Prodi_{s=0}^t { \lbc D^{A_w}(s) \rho^{A_w}(s \cond \bullet )ds  \rbc ^{ dA_w(s) }  \lbc D^{A_w}(s) \lp 1-\rho^{A_w}(s \cond \bullet )ds \rp  \rbc ^{ 1-dA_w(s) }}   \nonumber\\
    &=& \lsq \Prodi_{s=0}^t {\lbc D^{A_w}(s) \rho^{A_w}(s \cond \bullet )  \rbc ^{ \Delta A_w(s) }} \rsq \exp \lbc -\int_0^t  D^{A_w}(s) \rho^{A_w}(s \cond \bullet )ds \rbc,
\end{eqnarray}
where $\Delta A_w(t) = A_w(t)-A_w(t^-)$.  For individual $i$, both factors in~\eqref{eq:prod_int} need to be evaluated with respect to the individual's filtered counting process $\{N_{iq}^{A_w}(t): 0 \leq t \leq t_{K_i}, q =1, \ldots, Q_{w,i}\}$, 
where $Q_{w,i}$ is the number of initiations of treatment $w$ for individual $i$. Equation \eqref{eq:prod_int} then provides a basis for generalizing the discrete-time weights \eqref{eq:dt-ipw} to the continuous-time setting.

The first quantity in~\eqref{eq:prod_int} is equal to the finite product over the jump times and the second quantity is the survival function for treatment initiation. Thus, the continuous-time weight that corrects for potential selection bias associated with $A_w$ is the product of the density function $f^{A_w}$ and survival function $S^{A_w}$ of the filtered counting process for treatment $A_w$. In alignment with conventions established in prior literature \citep{hernan2001marginal,howe2012estimating}, the treatment weight for joint treatments $A_1$ and $A_2$ assumes a treatment order. By positing that treatment $A_1$ is administered infinitesimally earlier than treatment $A_2$, the intensity of initiating treatment $A_2$ at time $t$ can be dependent on the status of treatment $A_1$ at time $t$, $A_1(t)$, and the status of treatment $A_2$ at time $t^-$,  $A_2(t^-)$. Furthermore, the intensity of initiating treatment $A_1$ at time $t$ can rely on both $A_1(t^-)$ and $A_2(t^-)$. This assumption of ordered treatment administration will depend on specific clinical contexts. In Section~\ref{sec:sens_treat_order}, we carried out a sensitivity analysis to examine the impact of varying treatment order assumptions on the estimation of causal effects.  
Finally, based on \eqref{eq:prod_int}, the continuous-time stabilized weight for each individual is explicitly obtained as $\Omega^{A_1,A_2}(t_{K_i})= \Omega^{A_1}(t_{K_i})\Omega^{A_2}(t_{K_i})$, 
where
\vspace{-3pt}
\begin{align} 
\small
\begin{split} 
  & \Omega^{A_w}(t_{K_i}) \\
=&  \begin{cases}
     \dfrac{S^{A_w}\lp t_{K_i} \cond \bar{\mathcal{O}}^{A_w}(t_{K_i})\rp}{S^{A_w}\lp t_{K_i} \cond \bar{\mathcal{O}}_w(t_{K_i}) \rp} & \text{if $Q_{w,i}=0$}\\[3mm]
    \dfrac {f^{A_w}\lp U_{i,J_i-1} \cond \bar{\mathcal{O}}^{A_w}(U_{i,J_i-1}) \rp \lbc S^{A_w} \lp  V_{iJ_i} \cond \bar{\mathcal{O}}^{A_w}(V_{iJ_i}) \rp - S^{A_w}\lp t_{iK_i} \cond \bar{\mathcal{O}}^{A_w}(t_{K_i}) \rp \rbc} {f^{A_w}\lp U_{i,J_i-1} \cond \bar{\mathcal{O}}_w(U_{i,J_i-1}) \rp \lbc S^{A_w} \lp  V_{iJ_i} \cond \bar{\mathcal{O}}_w(V_{iJ_i}) \rp - S^{A_w}\lp t_{iK_i} \cond \bar{\mathcal{O}}_w(t_{K_i}) \rp \rbc }  & \text{if $Q_{w,i}=J_i-1$}\\[3mm]
    \dfrac{f^{A_w} \lp U_{iJ_i} \cond \bar{\mathcal{O}}^{A_w}(U_{iJ_i}) \rp }{ f^{A_w} \lp U_{iJ_i} \cond \bar{\mathcal{O}}_w(U_{iJ_i}) \rp} & \text{if $Q_{w,i}=J_i$},
    \end{cases}
    \end{split}
\end{align}
with $\bar{\mathcal{O}}_1(t) = \{\bar{A}_1(t^-), \bar{A}_2(t^-),  \bar{L} (t^-),T \geq t, C \geq t \}$, $\bar{\mathcal{O}}_2(t) = \{\bar{A}_1(t), \bar{A}_2(t^-),  \bar{L} (t^-),T \geq t, C \geq t \}$, $\bar{\mathcal{O}}^{A_1}(t) = \{\bar{A}_1(t^-), \bar{A}_2(t^-), T\geq t, C\geq t\}$, and $\bar{\mathcal{O}}^{A_2}(t) = \{\bar{A}_1(t), \bar{A}_2(t^-), T\geq t, C\geq t\}$. Supplementary Section 5 provides additional discussions of the weight formulation in connection to recurrent treatment initiations.
 
Turning to censoring, under the conditional exchangeability assumption (A2), the censoring process is covariate- and treatment-dependent. To correct for selection bias due to censoring, we additionally define a weight function associated with censoring, 
\begin{equation*}
    \Omega^C (G_i) = \frac{S^C \lp G_i \cond C_i \geq G_i, T_i \geq G_i \rp}{S^C \lp G_i \cond \bar{A}_1(G_i), \bar{A}_2(G_i), \bar{L}(G_i),  C_i \geq G_i, T_i \geq G_i\rp }, 
\end{equation*}
where $S^C$ is the survival function associated with the censoring process, and 
\begin{equation*}
G_i=\mathbbm{1}(\Delta_i^T =1)T_i+\mathbbm{1}(\Delta_i^T = 0,C_i > t_{K_i})t_{K_i}+
\mathbbm{1}(\Delta_i^T = 0,C_i \leq  t_{K_i} )C_i.
\end{equation*}
This leads to a modification of the estimating equation
for $\bm{\psi}$, 
\begin{equation} \label{eq:w_ee_final}
   \sum_{i=1}^n \int_0^\infty \Omega^{A_1,A_2} \Omega^C(G_i)  \lbc Z(A_{1i},A_{2i},t) - \bar{Z}^{\ast \ast}(t;\bm{\psi}) \rbc dN_i^T(t)=0, 
\end{equation} 
where 
$$\bar{Z}^{\ast\ast} = \dfrac{\sum_{k \in \mathcal{R}_t^T} Z(A_{k1}, A_{k2}, t) Y^{\ast\ast T }_{k}(t) r(A_{k1},A_{k2},t ;\bm{\psi})}{\sum_{k \in \mathcal{R}_t^T}Y^{\ast\ast T}_{k}(t) r(A_{k1},A_{k2},t ;\bm{\psi}) }$$ 
and 
$Y^{\ast  \ast T}_{i}(t) = \Omega^C(G_i) \Omega^{A_1,A_2}(t_{K_i}) Y^T_i(t)$. 

\subsection{Extensions to more than two time-varying treatments}\label{sec:general}
{While we present our methodology using two longitudinal treatments, our approach can be readily extended to accommodate multiple time-varying treatments in a straightforward manner. Theoretically, a fully interacted version of Model~\eqref{eq:JMSSM} can be developed to encompass all main effects of $a_w(t)$ for all $w \in \mathcal{W}$, along with their respective interactions. Considerations of clinical relevance and data sparsity for treatment combinations may further inform the inclusion of interaction terms within the structural model.  Suppose $\mathcal{B} = \{b_1(t), \ldots, b_{M}(t)\}$ is a collection of causal interaction effects of interest, 
e.g., $b_1(t) = a_1(t)a_2(t)$, the general joint marginal structural proportional hazards model is 
\begin{equation} \label{eq:JMSSM_general}
    \lambda^{T^{\jointa}} (t) = \lambda_0(t) \exp \lbc \sum_{w=1}^W\psi_{1w} a_w (t) + \sum_{m=1}^M \psi_{2m} b_m(t) \rbc,
\end{equation}
where the set of structural parameters, $\{\psi_{1w},w=1,\ldots,W\}$ and $\{\psi_{2m},m=1,\ldots,M\}$, capture the structural main and interaction effects on the counterfactual hazard function. A consistent estimator of $\bm{\psi}=\{\psi_{11},\ldots,\psi_{iW},\psi_{21},\ldots,\psi_{2M}\}$ can be obtained by solving the general form of the estimating equation
\begin{equation} 
   \sum_{i=1}^n \int_0^\infty \Omega^{A_1,\ldots,A_W} \Omega^C(G_i)  \lbc Z(A_{1i},\ldots, A_{Wi},t) - \bar{Z}^{\ast \ast}(t;\bm{\psi}) \rbc dN_i^T(t)=0, 
\end{equation} 
where $Z(A_{1i},\ldots, A_{Wi},t)$ is a vector of length $W+M$ representing observed time-varying treatments $A_w(t)$ and  multiplicative terms of those treatments whose causal interaction effects are of interest, and $\bar{Z}^{\ast \ast} $ is evaluated using the weighted risk set  $Y^{\ast \ast T}_{i}(t) = \Omega^C(G_i) \Omega^{A_1,\ldots,A_W}(t_{K_i})  Y_i^{T}(t)$. The estimation of joint treatment weights, denoted by $\Omega^{A_1, \ldots, A_W}(t_{K_i})$,  can be achieved by assuming predetermined sequence in which treatments are administered, as outlined  in Section \ref{sec:weights}. The estimation of  censoring weights $\Omega^C(G_i)$ adheres to the same methodology described in Section \ref{sec:weights}, except that the survival function in the denominator of $\Omega^C(G_i)$ is replaced by $S^C \lp G_i \cond \bar{A}_1(G_i), \bar{A}_2(G_i),\ldots,\bar{A}_W(G_i) ,\bar{L}(G_i),  C_i \geq G_i, T_i \geq G_i\rp$. 

\subsection{Estimation of the causal survival effects} \label{sec:weight-estimator}
We consider four ways in which the continuous-time treatment weights  can be estimated: (i) fitting a usual Cox regression model for the intensity process of the counting process of treatment initiation $\{A_w(t): t\in [0,t^o]\}$, estimating the density function $f^{A_w}$ and survival function $S^{A_w}$ from the fitted model with the Nelson-Aalen estimator for the baseline intensity, and finally calculating the weight $\Omega^{A_1,A_2}(t_{K_i})$ for each individual; (ii) smoothing the Nelson-Aalen estimator and in turn $f^{A_w}$ and $S^{A_w}$ estimated from the fitted Cox regression model by means of kernel functions \citep{ramlau1983smoothing}, and calculating the weights using the smoothed version of $f^{A_w}$ and $S^{A_w}$; (iii) fitting a multiplicative intensity tree-based model \citep{yao2022ensemble} in which the functional form of the intensity ratio for treatment initiation is flexibly captured to estimate the $f^{A_w}$ and $S^{A_w}$ and calculate the weights; (iv) smoothing the Nelson-Aalen estimator of the baseline intensity from the tree-based model and calculating the weights using the smoothed version of $f^{A_w}$ and $S^{A_w}$. Among these approaches, (i) relies on the parametric assumptions about the intensity ratio relationships between the treatment initiation process and covariate process and may be subject to model misspecification and bias for estimating causal effects. Compared to the Nelson-Aalen estimator which includes discrete jumps at event occurrences, the kernel function estimator in (ii) may help alleviate the issue of extreme or spiky weights, and has also been shown to be a consistent and asymptotically normal baseline intensity estimator \citep{andersen1993statistical}. Approach (iii) leverages a recent random survival forests model \citep{yao2022ensemble} that can accommodate time-varying covariates to mitigate the parametric assumptions and attendant biases associated with the usual Cox regression. In the context of baseline time-fixed treatments, previous research has used similar machine learning techniques to improve propensity score weighting estimators for both a point treatment \citep{lee2010improving,chang2022flexible} and a time-varying treatment \citep{shen2017estimation}. Additionally, these machine learning methods have been utilized to yield more accurate causal effect estimates in the presence of censored survival data \citep{hu2021estimating,hu2022cluster}. Finally, approach (iv) smooths the baseline intensity estimated from the survival forests for estimating the stabilized inverse probability weights, and serves as an additional step to smooth over the potentially spiky weights. In Section~\ref{sec:sim}, we compare the performances of these four strategies to estimating the continuous-time weights to generate practical recommendations. In addition, the censoring weight function $\Omega^C(G_i)$ can be estimated in a similar fashion via any one of these four approaches. Additional details of kernel function smoothing in approach (ii) and random survival forests in approach (iii) are presented in Supplementary Section 1. 

To accommodate the time-varying covariate process and account for the recurrent nature of treatment initiation, we fit a survival model to the counting process-stylized data. Each individual is represented by several rows of data corresponding to nonoverlapping time intervals of the form (start, stop]. To allow for discontinuous intervals of eligibility, when defining multiple time intervals $\mathcal{E}_{w,i} = \cup_{j=1}^{J_i} (V_{w,ij},U_{w,ij}]$ on $[0,t^o]$ for individual $i$, the duration of a treatment is removed from $[0,t^o]$ when the individual is currently being treated and therefore no longer eligible for initiating the treatment. Finally, since our estimators for $\bm{\psi}$ is a solution to the weighted partial score equation \eqref{eq:w_ee_final}, we can use the robust sandwich variance estimator to construct confidence intervals for the structural parameters; the details of the robust sandwich variance estimator is provided in Supplementary Section 1. Alternatively, nonparametric bootstrap can be used to construct confidence intervals that can take into account of the uncertainty of the continuous-time weight estimation. In practice, the robust sandwich variance estimator has been shown to be at most conservative under the discrete-time setting \citep{shu2021variance}, and we will empirically assess the accuracy of this variance estimator with continuous-time weights via simulations.

\section{Simulation Study}\label{sec:sim}

\subsection{Simulation design}
We carry out simulations to investigate the finite-sample properties of the proposed weighting estimators for the marginal structure Cox model parameters. We simulate data compatible with the marginal structural Cox model by generating and relating data adhering to the structural nested accelerated failure time (SNAFT) model \citep{young2008simulation}. 
A representation of a SNAFT model for time-varying treatment $a$ is \citep{hernan2005structural}
\[T^{\bar{0}} = \int_0^{T^{\bar{a}}} \exp\lsq \psi_{\text{aft}} a(t)\rsq, \]
where $T^{\bar{0}}$ is the counterfactual failure time under no treatment. 
This version of SNAFT assumes that both the left and right sides of the equation follow the same distribution. \cite{robins1992estimation} developed a simulation algorithm to generate data adhering to the SNAFT model under the discrete-time version of the identifying assumptions (A1)-(A3) in Section~\ref{sec:CT-IPW}. \cite{young2008simulation} showed that, under the same identifying assumptions, data adhering to a marginal structural Cox model of the  form $\lambda^{T^{\bar{a}}}(t) = \lambda_0(t) \exp \lsq \psi_{\text{msm}} a(t)  \rsq$
can be simulated from a SNAFT model with $\psi_{\text{aft}} = \psi_{\text{msm}}$ by adding an additional quantity to the term $\exp\lsq \psi_{\text{aft}} a(t)\rsq$. In particular, when $T^{\bar{0}}$ has an exponential distribution, the additional quantity is zero, hence the structural nested AFT model simulation algorithm \citep{robins1992estimation} can be used to appropriately simulate data compatible with the marginal structural Cox model under complex time-varying data structures. Building on these previous works, we extend the simulation algorithm described in \cite{karim2017estimating} to generate data from the joint marginal structural Cox model, while allowing for multiple time-varying treatments with discontinuous intervals of treatment eligibility and for both continuous and discrete time-varying confounders. 

Throughout we simulate an observational study with $n=1000$ patients and two time-varying treatments $A_1(t)$ and $A_2(t)$. We assume $\bar{L}(t)$ is appropriately summarized by a continuous time-varying confounding variable $L_1(t)$ and a binary time-varying confounding variable $L_2(t)$. The simulation algorithm includes two steps. In step (1), we consider nonlinear terms for the continuous variable $L_1(t_k)$ and the interaction term $A_1(t_{k-1}) \times L_1(t_k)$, $A_2(t_{k}) \times L_1(t_k)$,  $A_1(t_{k-1}) \times L_2(t_k)$ and $A_2(t_{k}) \times L_2(t_k)$ in the true treatment decision model. In particular, past treatment status $\{A_1(t_{k-1}), A_2(t_k)\}$ is a predictor of $L(t_k)$, which then predicts future treatment exposure $\{A_1(t_{k}), A_2(t_{k+1}) \}$ as well as future failure status $Y(t_{k+1})$ via $1/ \log(T^{\bar{0}}$). Therefore, $L(t_k)$ is a time-dependent confounder affecting both the future treatment choices and counterfactual survival outcomes. The simulation of treatment initiation is placed in the recurrent event framework. Once treatment is initiated at time $t_k$, treatment duration following initiation is simulated from a zero-truncated Poisson distribution. In step (1), we generate a longitudinal data set with $100\times 1000$ observations (100 aligned measurement time points for each of $n=1000$ individuals). In step (2), we randomly discard a proportion of follow-up observations for a randomly selected subset of individuals \citep{lin2004analysis}; and in the resulting data set, the individuals will have varying number of follow-up measurement time points, which are also irregularly spaced.  Supplementary Section 2 provides the full pseudo-code for simulating data under the marginal structural Cox model with two time-varying treatments. To evaluate the performance characteristics of the proposed method with respect to sample size, we additionally examine five smaller sample sizes, $n=250$, $n=350$, $n=500$, $n=650$ and $n=850$, all featuring 100 follow-up time points for each individual. 

Our simulation parameters are chosen so that the simulated data possess similar characteristics to those observed in the motivating COVID-19 data set. The treatments $A_1$ and $A_2$ are simulated to resemble dexamethasone and remdesivir such that: (i) about 20\% patients did not take any of the anti-viral and anti-inflammatory medications aimed at treating COVID-19; (ii) among those who were treated, 62\% took dexamethasone only, 25\% took remdesivir only and 13\% took both (either concurrently or with treatment switching); (iii) the number of initiations for both treatments ranges from 0 to 4 with the average medication duration about 5 days. The values of treatment effect parameters $\psi_1$ and $\psi_2$ were set to yield a 6.7\% mortality proportion among those who received dexamethasone and a 4.9\% mortality proportion in those treated with remedesivir. 

\subsection{Comparison of methods}
We conduct two sets of simulations to investigate the finite-sample performance of our proposed joint marginal structural survival model in continuous time  (JMSSM-CT). First, we compare how accurate the four weighting estimators described in Section~\ref{sec:weight-estimator} estimate the structural parameter $\bm{\psi}$. For ease of comparison, in this simulation we consider the structural parameter $\bm{\psi}$ as the target of inference. Because the counterfactual survival functions directly depend on the structural parameter $\bm{\psi}$, unbiased estimation of $\bm{\psi}$ will lead to improved estimation of counterfactual survival functions or any summaries of these counterfactual quantities (such as median counterfactual survival time or restricted mean survival time). Second, we use the best weighting estimator, suggested by the first set of simulation, for JMSSM-CT, and compare it with the joint marginal structural model that requires aligned discrete time points (JMSM-DT). To ensure an objective comparison, we use the random forests \citep{yao2022ensemble} and adapt it into our proposed recurrent event framework to estimate the weights for JMSM-DT. In addition, we implement both JMSSM-CT and JMSM-DT on the ``rectangular'' simulation data with 100 aligned time points for each individual and on the ``ragged'' data with irregular observational time points. The performance on the rectangular data will be considered as the benchmark performance, based on which we will assess the relative accuracy of JMSSM-CT and JMSM-DT when estimating the structural parameters with the ``ragged'' data.  

\subsection{Performance characteristics}\label{sec:perform}

\begin{figure}[htbp]
    \centering
    \includegraphics[width=0.85\textwidth]{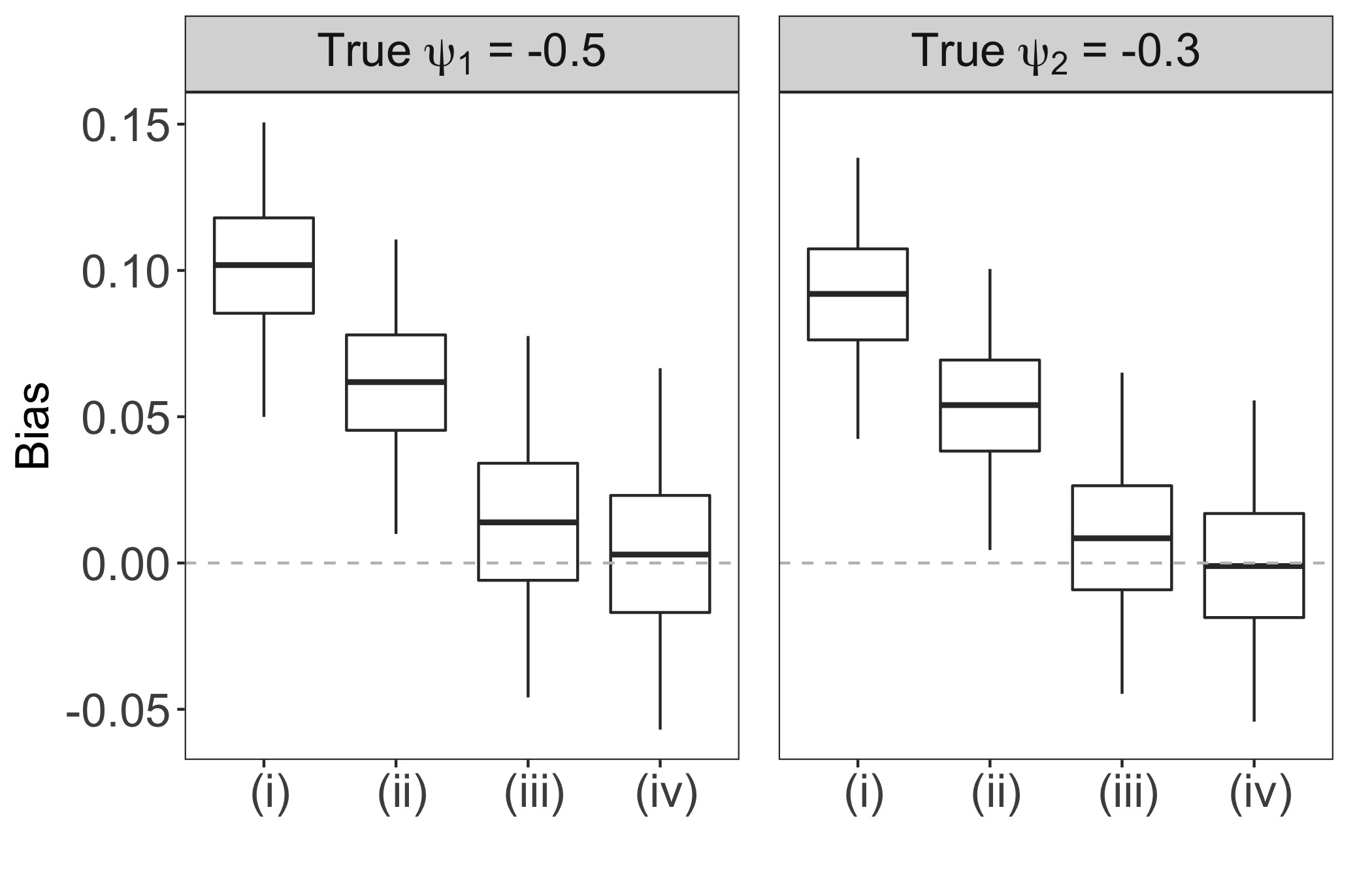}
    \caption{Biases in estimates of $\psi_1$ and $\psi_2$ for sample size $n=1000$ across 250 data replications using four approaches to estimate the weights as described in Section~\ref{sec:weight-estimator}. Approach (i) uses main-effects Cox regression model and Nelson-Aalen estimator for baseline intensity. Approach (ii) uses kernel function smoothing of the Nelson-Aalen estimator in approach (i). Approach (iii) uses a survival forests model that accommodates time-varying covariates and Nelson-Aalen estimator for baseline intensity. Approach (iv) uses kernel function smoothing of the Nelson-Aalen estimator in approach (iii).}
    \label{fig:bias_4weight_estimator}
\end{figure}

To assess the performance of each method, we simulate 250 observational data sets using the above approach, and evaluate the absolute bias, root mean squared error (RMSE) and covarage probability (CP) for estimating the $\bm{\psi}$. The CP is evaluated on normality-based confidence intervals with the robust sandwich variance estimator. Additionally, we empirically evaluate the convergence rate of the proposed method by measuring how rapidly the RMSE decreases as the sample size increases \citep{hu2020estimation}. 
 
In Figure~\ref{fig:bias_4weight_estimator}, we compared the biases in estimating $\psi_1$ and $\psi_2$ for a sample size of $n=1000$ by employing the four weight estimation approaches. The weighting estimator (iv) using both the flexible tree-based survival model and kernel function estimator of the treatment initiation intensity yielded the lowest biases in estimating both $\psi_1$ and $\psi_2$. By contrast, the weighting estimator (i) via the usual main-effects Cox regression model, susceptible to model misspecification, along with the Nelson-Aalen baseline intensity estimator produced the largest estimation bias. Applying the kernel function smoothing to the Nelson-Aalen estimator led to bias reduction for both the Cox (approach (ii)) and tree-based survival model (approach (iv)) for the treatment process. Flexible modeling of the intensity ratio function  has a larger effect in reducing the bias in structural parameter estimates than smoothing the nonparametric baseline intensity estimator. For example, compared to approach (ii), approach (iii) further reduced the mean absolute bias (MAB) in estimating $\hat{\psi}_1$ by approximately 67\%. 
Supplementary Table 1 summarizes the MAB, RMSE and CP for the four weighting estimators and similarly suggests that approach (iv) led to the smallest MAB and RSME, and provided close to nominal CP with the robust sandwich variance estimator. 

Supplementary Figure 1 presents biases in estimating $\psi_1$ and $\psi_2$ for sample sizes $n=$ 250, 350, 500, 650, 850, and 1000 across 250 replications. These results suggest that the biases decrease for all weighting estimators as the sample size increases. Supplementary Figure 2 illustrates the approximate relationship between RMSE and the sample size $n$. By performing a simple linear regression of $\log (\text{RMSE})$ on ($-\log n$) and obtaining the slope $b$ for each weighting estimator, we assessed the approximate convergence rate of each weighting estimator through the least-squares estimation of $b$. Estimators (iii) and (iv) approximately converged at a faster rate of $O(n^{-1/2})$, while estimators (i) and (ii) approximately converged more slowly at $O(n^{-1/3})$.

Suggested by an anonymous reviewer, we additionally implemented  JMSSM-CT utilizing a weighting estimator that focused on estimating the effect of a single treatment, while treating the other treatment as a time-varying confounder. We then compared these results to those obtained when employing joint treatment weights, as presented in Supplementary Figure 3. Upon estimating both structural parameters $\psi_1$ and $\psi_2$, it became evident that each of the four weighting estimators utilizing joint treatment weights produced significantly smaller biases in comparison to the estimators that were dedicated to estimating the effects of a single treatment.

The second set of simulation benchmarks the performance of JMSSM-CT versus JMSM-DT on the data with fully aligned follow-up time points and compare how much each method can recover the benchmark performance in situations where the longitudinal measurements are irregularly spaced. Table~\ref{tab:sim_CTvsDT} displays the MAB, RMSE and CP for each of the two methods under both data settings, and Supplementary Figure 4 visualizes the distributions of biases across 250 data replications. In the rectangular data setting with fully aligned time points, compared to JMSM-DT, JMSSM-CT had similar CP but smaller MAB and RMSE.  As the sparsity of longitudinal measurements increased and the time intervals became unevenly spaced, the proposed JMSSM-CT could still recover the benchmark performance; whereas the JMSM-DT had a deteriorating performance (larger MAB and RMSE and lower CP), with larger performance decline under coarser discretization of the follow-up time. Supplementary Table 2 summarizes the distribution of estimated individual time-varying weights from one random replication of the ragged data for JMSSM-CT and JMSM-DT. Overall, JMSSM-CT with the weighting estimator (iv) provided the smallest maximum/minimum weight ratio (2.36 / 0.68) and no extreme or spiky weights. 

\begin{table}[htbp]
\caption{Comparing the proposed method JMSSM-CT in continuous time with JMSM-DT in discrete time in estimating the treatment effect $\bm{\psi}$ on the bases of mean absolute bias (MAB), root mean square error (RMSE) and coverage probability (CP) across 250 data replications. In the estimation of the weights, the weighting estimator (iv) was used for JMSSM-CT and the random forests adapted into our recurrent event framework (Section 3.2) was used for JMSM-DT. Both methods were implemented on the ``rectangular'' simulation data with 100 aligned time points for each individual and on the ``ragged'' data with unaligned time points. With the ragged data, the follow-up time was discretized in the space of 0.5 , 1 and 2 days for JMSM-DT.} 
\def\arraystretch{1.2}
\begin{tabular}{llccccccccccc}\\
\toprule
\multirow{2}{*}{Data format} & \multirow{2}{*}{Methods} &\multicolumn{3}{c}{$\psi_1$} && \multicolumn{3}{c}{$\psi_2$}\\
\cline{3-5} \cline{7-9}
 && MAB & RMSE & CP && MAB & RMSE & CP  \\
\midrule
\multirow{2}{*}{Rectangular} & JMSM-DT & .021 & .026   & .944 && .019 &.023  &  .948 \\ 
& JMSSM-CT & .015 &.020  &  .948 && .014& .018 &   .948 \\ 

\multirow{4}{*}{Ragged}&JMSM-DT (2d) & .040 & .047   & .660 && .035 & .041   &  .668 \\ 
&JMSM-DT (1d) & .033 & .041   & .732 && .029 & .035   &  .738 \\ 
&JMSM-DT (0.5d) & .027 & .034   & .801 && .024 & .030   &  .804 \\ 
& JMSSM-CT & .016 &.022  &  .952 && .015& .019 &   .952 \\ 

\bottomrule
\end{tabular}
\label{tab:sim_CTvsDT}
\end{table}

All simulations were executed in the \texttt{R} programming environment on an iMac equipped with a 4 GHz Intel Core i7 processor. On a dataset comprising 100 aligned time points for each of the 1,000 individuals ($n = 1000$), the execution times for each weighting estimator per data replication in the simulation were as follows: (i) 0.35 minutes , (ii) 0.64 minutes , (iii) 8.12 minutes , and (iv) 8.45 minutes.

\section{Estimating Causal Effects of Multiple COVID-19 Treatments}\label{sec:analysis}
\subsection{Description of the COVID-19 data}\label{sec:data}
We apply the proposed method JMSSM-CT to a comprehensive COVID-19 data set drawn from the Epic electronic medical records system of the Mount Sinai Medical Center, and draw causal inferences about the comparative effectiveness of multiple COVID-19 treatments. The data set includes 11,286 de-identified unique adult patients ($\geq$18 years of age) who were diagnosed with COVID-19 and hospitalized within the Mount Sinai Health System between February 25, 2020 to February 26, 2021. A confirmed case of COVID-19 was defined as a positive test result from a real-time reverse-transcriptase PCR-based clinical test carried out on nasopharyngeal swab specimens collected from the patient \citep{wang2020hospitalised}.

We focus on the comparative effectiveness of four treatment classes that are of most clinical interest: (i) remdesivir; (ii) dexamethasone; (iii) anti-inflammatory medications other than corticosteroids; and (iv) corticosteroids other than dexamethasone. We defined treatment classes by carefully reviewing the medications administered to patients. For example, the dexamethasone class includes both oral and intravenous dexamethasone; and the corticosteroids other than dexamethasone class includes oral and intravenous hydrocortisone, oral and intravenous methylprednisolone, intravenous prednisolone, and oral and intravenous prednisone.  Detailed definitions of the four treatment classes are provided in Supplementary Table 3. The observed treatment patterns are visualized in Figure~\ref{fig:treat_pattern};  patients could be simultaneously taking two or more treatment classes, or they could switch from one treatment class to another during their hospital stays. Treatment alterations or discontinuations may be influenced by indications of therapeutic failure, such as deteriorating oxygen saturation levels, adverse side effects like bleeding or disseminated intravascular coagulation, or evidence of therapeutic effectiveness, such as improved oxygen saturation levels.

Following suggestions by our clinician investigators, we assumed that the following time-fixed and time-varying confounders were sufficient to predict both treatment decision and outcome (i.e., based on which assumption (A2) ic considered plausible):  age, sex, race, ethnicity, D-dimer levels (the degradation product of crosslinked fibrin, reflecting ongoing activation of the hemostatic system),  serum creatinine levels (a waste product that forms when creatine breaks down, indicating how well kidneys are working), whether the patient used tobacco at the time of admission, history of comorbidity represented by a set of binary variables: hypertension, coronary artery disease, cancer, diabetes,  asthma and chronic obstructive pulmonary disease, hospital site, and patient oxygen levels (definition provided in Supplementary Table 4). The time-varying confounding variables were D-dimer levels, serum creatinine level and patient oxygen levels.

\begin{figure}[h!b!p!]
    \centering
    \includegraphics[width=1\textwidth]{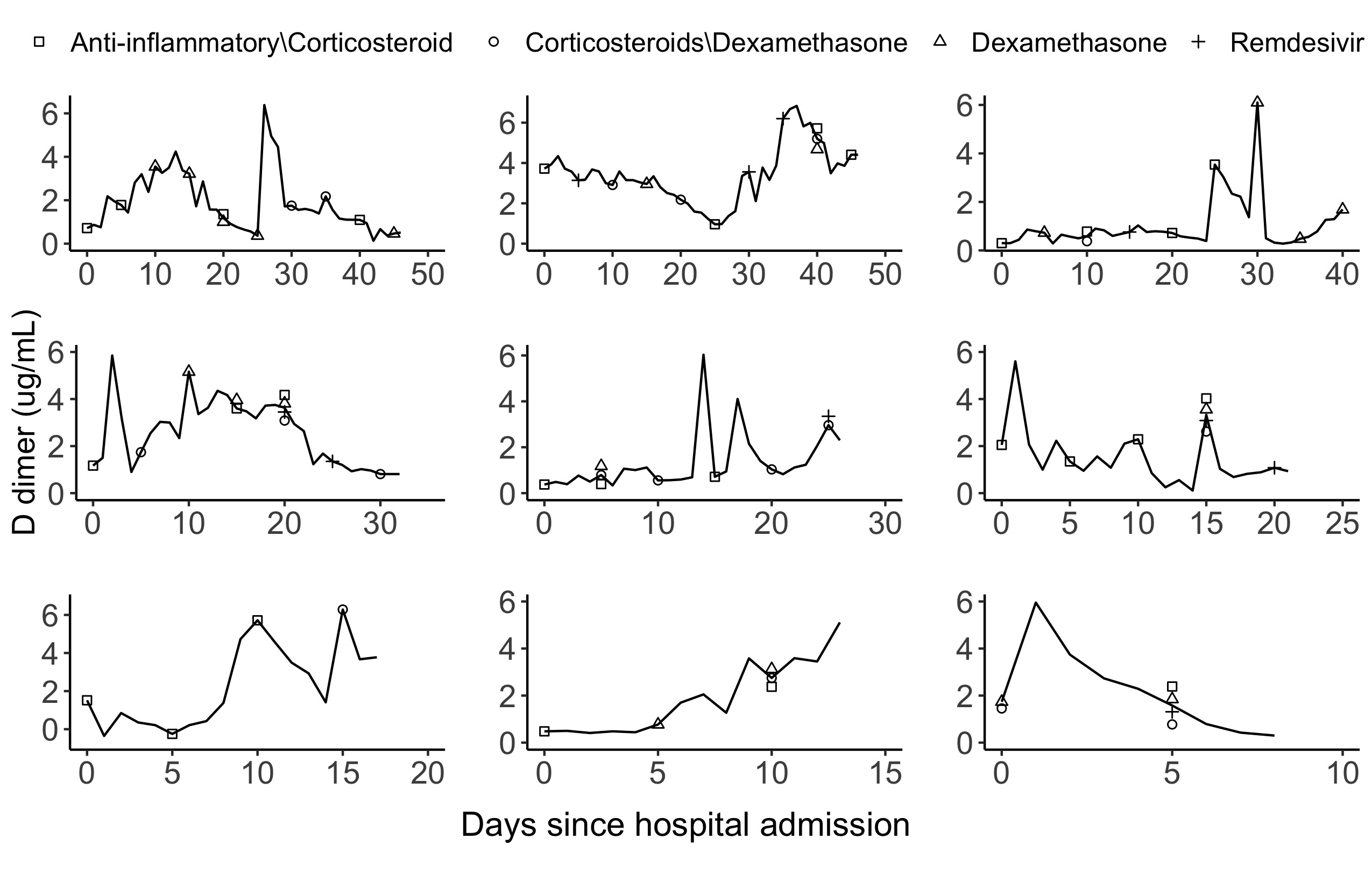}
    \caption{Trajectories of D-dimer levels over the course of hospital stay for 9 randomly chosen patients. Symbols represent the types of treatment classes received by a patient at a given time.}
    \label{fig:ddimer}
\end{figure}

The average age of this sample population is 64.6 with a standard deviation of 18.1. About 54\% of the patients were male and 46\% female. The Hispanics accounted for about 26\% of the patient population and the racial composition is 29\% Whites, 25\% Blacks, 6\% Asians and 40\% Other. Summary statistics for time-fixed confounders are presented in Supplementary Table 5. Time-varying confounders were measured repeatedly over the course of hospital stay. Figure~\ref{fig:ddimer} displays  trajectories of  D-dimer levels for 9 randomly chosen patients over the course of hospital stay. Supplementary Figure 5-6 show trajectories of the serum creatinine levels and patient oxygen levels. A considerable variability is observed across patients in both these time-varying measures and treatments. 

We considered a composite outcome, ICU admission or in-hospital death, whichever occurs first. The outcome may be right censored by hospital discharge or administratively censored on $t^o$= February 26, 2021, the date on which the database for the current analysis was locked. 

\subsection{Marginal structural modeling, assumptions and estimands}\label{sec:mod-assump-est}
We implemented all four approaches discussed in Section~\ref{sec:weight-estimator} to estimate the time-varying weights for the JMSSM-CT model. Additionally, we discretized the time in the space of 1, 3 and 5 days, and applied the discrete time based method JMSM-DT to compare with our proposed JMSSM-CT method. The weight model included all time-fixed and time-varying confounders listed in Supplementary Table 5 and shown in Figure~\ref{fig:ddimer} and Supplementary Figure 5-6. No variable selection and no nonlinear transformations of the confounders were performed prior to model fitting.  When fitting the joint marginal structural proportional hazards model~\eqref{eq:JMSSM_general}, pairwise treatment interactions were included if there were sufficient data points supporting the joint use of the pair of treatments. 

Of note, the JMSSM-CT model should be interpreted under several assumptions, some of which are structural and may not be verifiable from the observed data alone. Firstly, the consistency assumption (A1) requires that the observed time to composite outcome or censoring time corresponds to their counterfactual values under a specific joint treatment trajectory, and rules out patient-level interference. This is considered plausible in the current application, because specialized COVID-19 medications are rare and that a patient's survival outcome generally is not affected by medication trajectories of other patients even in the same ward. Likewise, hospital discharge has also been carefully reviewed by physicians, and thus not affected by other patients' medication plan. Secondly, the fixed and time-varying confounders described in Section \ref{sec:data} are carefully chosen after discussions with the investigator team and reflect the current understanding about the COVID-19 treatment decisions. Therefore, the conditional exchangeability assumption (A2) is considered plausible. Violations of this assumption can arise, for example, if the medication has been provided purely based on physician preference which is generally unmeasurable. In these cases, sensitivity analyses with posited sensitivity parameters would be helpful for results interpretation, even though development of such methodology in continuous-time is currently limited and requires future work. Thirdly, we indirectly assessed the positivity assumption (A3) by visualizing the estimated individual time-varying weights in Table \ref{tab:weight_dist} and Figure \ref{fig:tv-weight} (details in Section \ref{sec:res}). No extreme weights were identified, and there has been no strong indication against positivity.

Besides the aforementioned structural assumptions, the fitted JMSSM-CT model itself includes proportional hazards as a modeling assumption. Therefore, our primary analysis should be interpreted under the condition that the marginal structural Cox model is correctly specified. When this condition holds, the structural regression parameter represents a causal hazard ratio, as this is precisely the case where the hazard ratio parameter can be written as a ratio between two log counterfactual survival functions. However, we acknowledge that more generally a time-varying hazard ratio parameter lacks causal interpretation due to built-in selection bias issues; see \citet{hernan2010hazards} and \citet{martinussen2020subtleties} for a detailed treatment on this issue. Recognizing the possible limitations of the proportional hazards assumption in parameterizing the treatment effects for analyzing the COVID-19 data, we additionally fit the marginal structural additive hazards model in continuous time (see Section~\ref{sec:res}). In operational terms, the only modification involves replacing the weighted partial score equations in \eqref{eq:w_ee} with the weighted least squares equations under the additive risk model of \citet{lin1994semiparametric}. To facilitate the comparison between these models, we focus on interpreting the counterfactual survival functions as well as counterfactual restricted mean survival times under specific treatment regimens. Finally, while the continuous-time weights generally require some modeling assumptions, the application of flexible tree-based survival model can partially alleviate bias due to parametric modeling assumptions (for the treatment initiation intensity ratio function), as demonstrated in Section \ref{sec:perform}.  The 95\% confidence intervals for all structural regression parameters are obtained based on the robust sandwich variance estimators. However, for estimating the estimands based on the counterfactual survival functions and restricted mean survival times, we consider bootstrap-based confidence interval estimators with $1000$ bootstrap replicates.

\subsection{Results}\label{sec:res}
Using the stabilized inverse probability weights to correct for time-varying confounding and censoring, the structural model parameter estimates $\hat{\bm{\psi}}$ (log hazard ratio) and the associated 95\% confidence intervals are provided in Table \ref{tab:param_est_covid}. Echoing the findings from our simulation study (Section~\ref{sec:sim}), the weighting estimator (iv), using the random survival forests model and kernel function smoothing of the Nelson-Aalen estimator, produced the narrowest confidence intervals. By contrast, the weighting estimator (i), using the main-effects Cox regression model and non-smoothed Nelson-Aalen estimator, led to the widest confidence intervals. As a result, using the weighting estimator (iv), we observe  a statistically significant treatment benefit (in log hazard ratio) with  dexamethasone ($-.2 (-.35, -.06)$) and remedesivir ($-.53 (-.75, -.31)$), and added treatment benefit if remedesivir is used in combination with corticosteroids other than dexamethasone. Using the weighting estimator (i), none of the main or interactive treatment effects appeared to be statistically significant. 

\begin{table}[H]
\centering
\caption{The joint and interactive effect estimates $\hat{\bm{\psi}}$ (log hazard ratio) of COVID-19 treatments and associated 95\% confidence intervals (CI), using the COVID-19 dataset drawn from  the  Epic  electronic  medical  records  system  at  the  Mount  Sinai  Medical  Center. The composite outcome of in-hospital death or admission to ICU was used. To estimate the weights, four approaches (i)-(iv) (Section~\ref{sec:weight-estimator}) were used for JMSSM-CT. Confidence intervals were estimated via the robust sandwich variance estimators. ``$\times$" denotes treatment interaction.}
\vspace{0.1in}
\def\arraystretch{1.1}
\setlength{\tabcolsep}{1pt}
\scalebox{0.9}{
\hspace*{-3em}
\begin{tabular}{lcccc}
\toprule
\multirow{2}{*}{Treatment classes}&\multicolumn{4}{c}{ $\hat{\bm{\psi}}$ (95\% Confidence Interval)} \\
\cline{2-5}
  & (i) & (ii) & (iii) & (iv)\\
\midrule
Dexamethasone & $-0.02(-0.45, 0.41)$ & $-0.15 (-0.36, 0.06)$ &   $-0.19(-0.36, -0.02)$& $-0.20 (-0.35, -0.06)$\\\smallskip 

Remdesivir  & $-0.22(-0.61, 0.16)$ & $-0.48(-0.76, -0.20)  $&   $-0.55(-0.78, -0.32) $& $-0.53 (-0.75, -0.31)$\\
Corticosteroids other than & & &&\\\smallskip 
 dexamethasone & $0.14(-0.21, 0.49) $ &$-0.02(-0.35, 0.31) $ &  $-0.06(-0.34, 0.24) $ &  $-0.08 (-0.29, 0.19)$\\
Anti-inflammatory medications & & &&\\\smallskip 
other than corticosteroids & $0.15(-0.41, 0.72)$ &$0.01(-0.52, 0.54)$  &   $-0.03(-0.62, 0.56)$& $-0.05 (-0.56, 0.47)$\\
Remdesivir $\times$ Corticosteroids & & &&\\\smallskip 
other than dexamethasone & $-0.22(-0.68, 0.24)$&$-0.67(-0.94, -0.40)$  &   $-0.69(-0.96, -0.42)$&  $-0.74 (-0.95, -0.52)$\\ 
\bottomrule
\end{tabular}}
\label{tab:param_est_covid}
\end{table}

To obtain further insights into the operating characteristics of each method, we summarize the distribution of the estimated individual weights in Table~\ref{tab:weight_dist}. As one can clearly see, the weighting estimator (i) produced a substantial amount of extreme weights -- the minimum of .0001 and maximum of 63. By comparison, the estimator (iv) generated no spiky weights, with the mean value of close to one. There is little difference in the weight distribution between estimator (ii) and estimator (iii), both of which mitigated the issue of extreme weights, but not to the same degree as the estimator (iv).  Figure~\ref{fig:tv-weight} shows the side-by-side comparison of the time-varying weights at 7, 14, 21 and 28 days since hospital admission estimated using the four weighting estimators. The weighting estimator (iv) produced no extreme weights at any of the time points. An increasing amount of extreme weights was generated when the modeling flexibility decreased or when the baseline intensity estimator was not smoothed.

\begin{table}[H]
\centering
\caption{The distribution of the individual time-varying weights estimated from the COVID-19 data. Four approahces (i)-(iv) described in Section~\ref{sec:weight-estimator}  were used for the weight estimation.}
\begin{tabular}{ccccccccc}
\toprule
\multirow{2}{*}{weighting estimators}&\multicolumn{5}{c}{Distribution of estimated weights}\\ \cline{2-6}
& Minimum & First quartile & Mean & Third quartile & Maximum   \\
\midrule
(i) & 0.0001 & 0.172   & 4.443 & 3.367 & 63.112   \\ 
(ii) & 0.004 & 0.132   & 0.915 & 1.308 & 4.018   \\ 
(iii) & 0.004 & 0.154   & 1.096 & 1.511& 5.260   \\ 
(iv) & 0.088 & 0.340   & 0.957 & 1.367 & 2.895   \\ 
\bottomrule
\end{tabular}
\label{tab:weight_dist}
\end{table}

\begin{figure}[H]
    \centering
    \includegraphics[width=.75\textwidth]{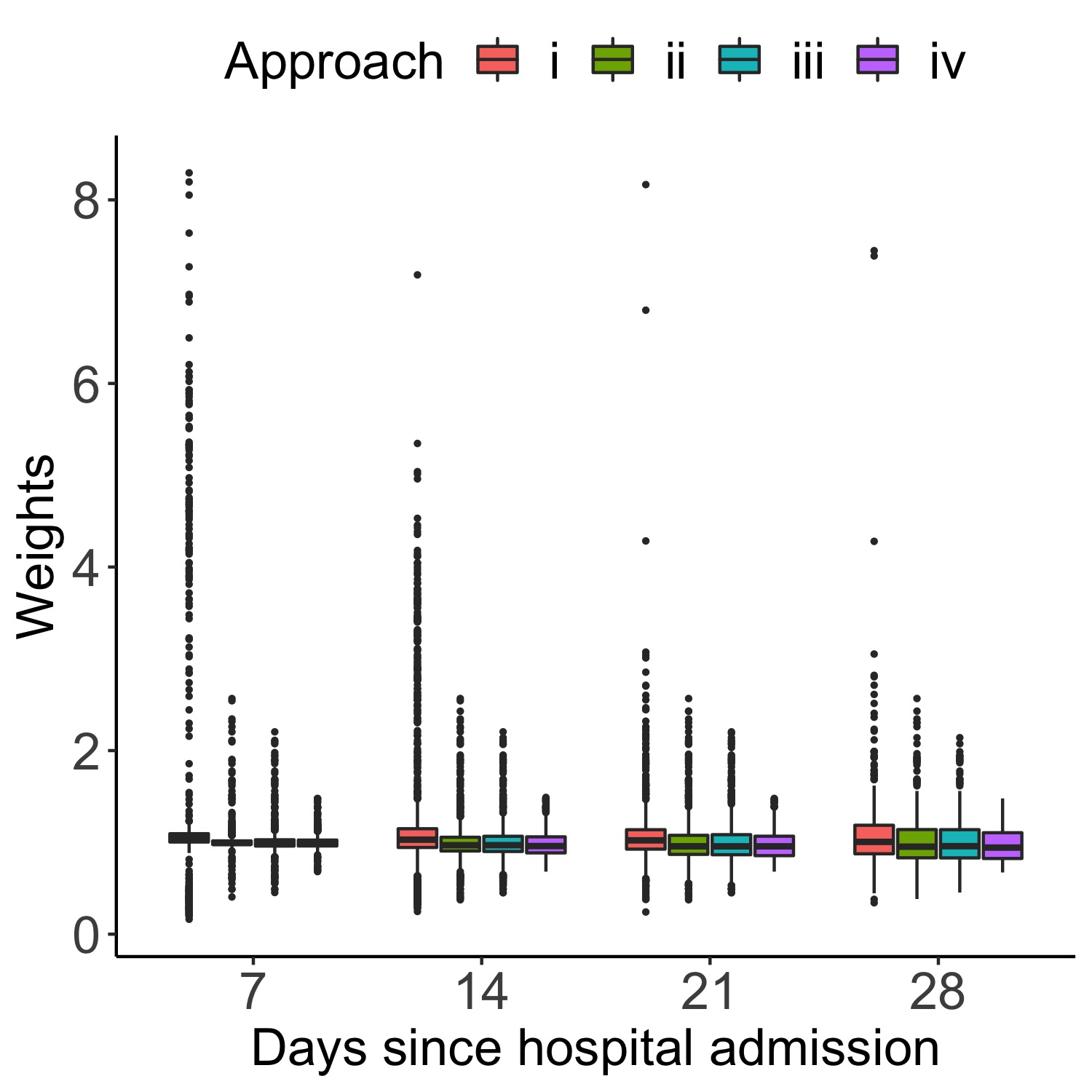}
    \caption{Side-by-side comparison of the time-varying weights estimated via approach (i)-(iv) described in Section~\ref{sec:weight-estimator}.}
    \label{fig:tv-weight}
\end{figure}

Corroborating findings from our simulation study, discretizing the time can lead to the loss of information and efficiency, suggested by the  width of confidence intervals of $\bm{\psi}$ for JMSSM-CT and JMSM-DT, shown in Supplementary Table 6. The JMSSM-CT yielded the narrowest confidence intervals; whereas for JMSM-DT, the width of confidence intervals grows as the space of days for the discretization increases.

 Figure~\ref{fig:Surv_overall} displays the counterfactual survival curves for ICU admission or death (whichever occurs first) among COVID-19 patients, calculated using the $\hat{\bm{\psi}}$ estimated from the most promising weighting estimator (iv). These curves represent the outcomes under five distinct treatments administered upon hospital admission and maintained throughout the patients' hospital stays. Among the four main treatment classes, remdesivir had significantly better treatment benefits followed by dexamethasone than two alternative treatment classes: anti-inflammatory medications other than corticosteroid and corticosteroids other than dexamethasone.  Interestingly, remdesivir and corticosteroids other than dexamethasone had a significant treatment interaction effect suggesting additional survival benefit when they are used in combination. This is demonstrated by the highest counterfactual survival curve under the concomitant use of these two types of medications. Table~\ref{tab: survprob-rmst-covid} presents the 14-day counterfactual survival probabilities and restricted mean survival times for the five distinct treatments; the results are consistent with the findings illustrated in Figure~\ref{fig:Surv_overall}. Recognizing the potential limitation of the proportional hazards assumption in our joint marginal structural survival model, we performed an additional comparative effectiveness analysis for the five treatments using the joint marginal structural additive hazards model. The counterfactual survival probabilities and restricted mean survival times obtained from this analysis are presented in Supplementary Table 7, which show a high degree of similarity to the results displayed in Table~\ref{tab: survprob-rmst-covid} using our marginal structural portional hazards model.

\begin{table}[H]
\centering
\caption{14-day counterfactual survival probability and restricted mean survival time (RMST) using the proposed method JMSSM-CT with weighting estimator (iv). The composite outcome of in-hospital death or admission to ICU was used. The 95\% confidence intervals were computed using nonparametric bootstrap with 100 replications. The symbol ``$\smallsetminus$'' indicates ``other than''. } 
\vspace{0.1in}
\def\arraystretch{1.1}
\hspace*{-3em}
\begin{tabular}{p{8.5cm}cccc}
\toprule
Treatment &  Survival probability (14d) & 14-day  RMST\\ 
\midrule
Dexamethasone &  0.890 (0.873, 0.906) & 13.68 (13.43, 14.03)\\ 
Remdesivir  &  0.924 (0.910, 0.938) & 13.79 (13.55, 14.05)\\
Corticosteroids$\smallsetminus$dexamethasone & 0.861 (0.845, 0.877) & 13.29 (13.05, 13.53) \\
Anti-inflammatory medications $\smallsetminus$ corticosteroids & 0.843 (0.827, 0.859) & 13.36 (13.11, 13.61)\\
Remdesivir + Corticosteroids 
$\smallsetminus$ dexamethasone &  0.941 (0.928, 0.955) & 13.85 (13.64, 14.09)\\
\bottomrule 
\end{tabular}
\label{tab: survprob-rmst-covid}
\end{table}

\begin{figure}[!htbp]
    \centering
    \includegraphics[width=0.9\textwidth]{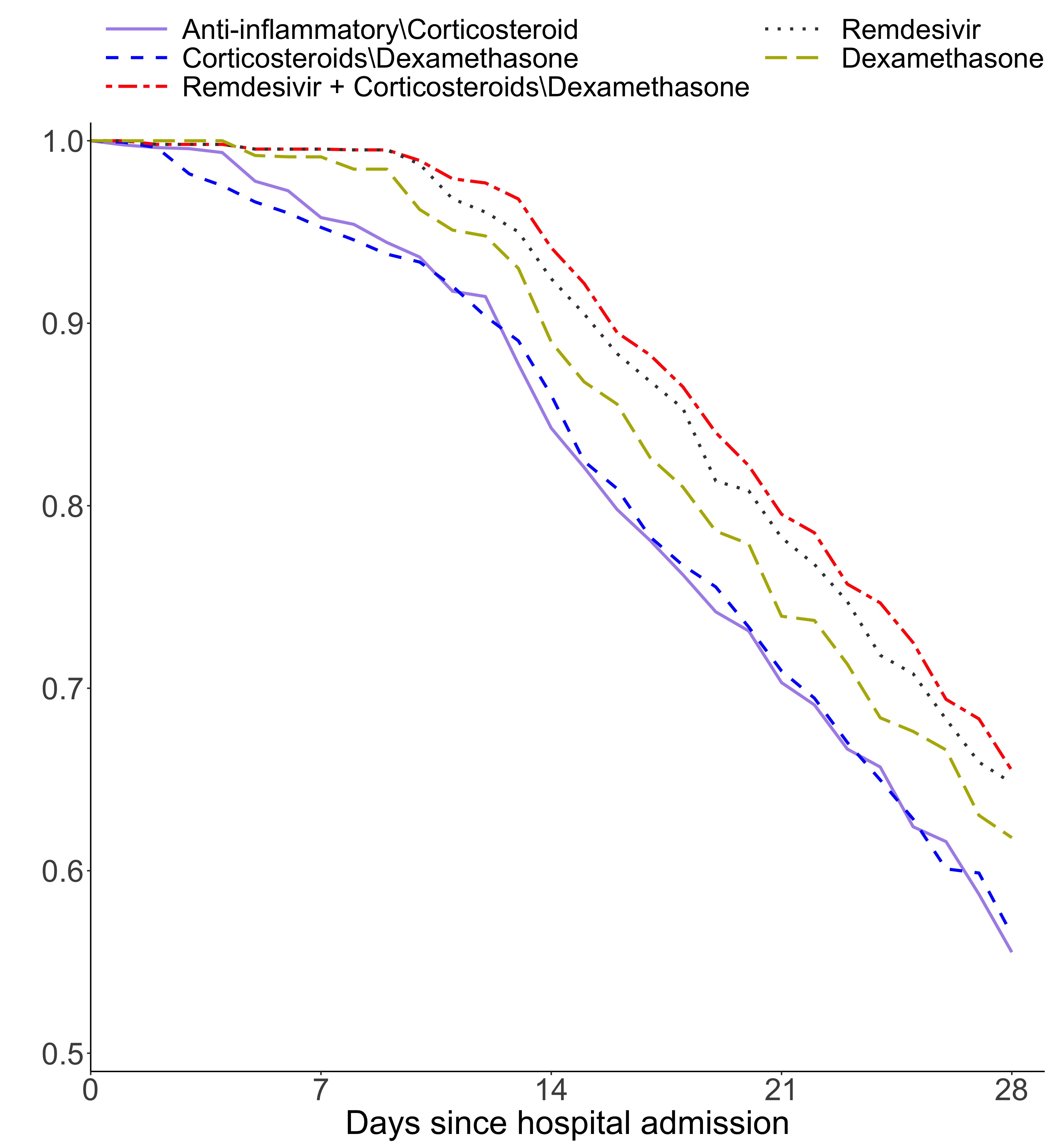}
    \caption{Counterfactual survival curves for each of five treatments among the general COVID-19 patients. The composite outcome of ICU admission or death is used. }
    \label{fig:Surv_overall}
\end{figure}

\subsection{Sensitivity analysis to assess the impact of treatment ordering} \label{sec:sens_treat_order}
When multiple time-varying treatments are under investigation, the analyst may face the question of the optimal treatment ordering by which the joint treatment weights $\Omega^{A_1, \ldots, A_W} (t_{K_i})$ (Section~\ref{sec:general}) can be estimated. In our simulation and case study, we used the default decreasing order of ``treatment sizes''. Note that the ``treatment sizes'' considered here do not refer to the sizes of mutually exclusive treatment groups, but simply the number of patients who have received the treatments at some point during the hospital stay. In our COVID-19 dataset, 7830 patients took dexamethasone ($A_1$) at some points in time following hospital admission, 4943 took corticosteroids other than dexamethasone ($A_2$),  4103 received remedesivir ($A_3$), and 2844 were treated with anti-inflammatory other than corticosteroid ($A_4$).  We estimated the joint treatment weights by positing that  $A_1$ is administered infinitesimally earlier, followed sequentially by $A_2$, $A_3$ and $A_4$. This implies that upon estimating the joint treatment weights, the intensity of initiating treatment $A_1$ at time $t$ can depend on $A_1(t^-)$, $A_2(t^-)$,  $A_3(t^-)$ and $A_4(t^-)$. By comparison, the intensity of initiating treatment $A_w$ can be conditional on $A_w(t^-)$ and $A_1(t), \ldots, A_{w-1}(t)$, for $2 \leq w \leq 4$.

To evaluate whether treatment ordering can impact the causal inferences about treatment effects, we conducted a sensitivity analysis, in which we explored four choices of treatment order: decreasing order of treatment sizes, increasing order of treatment sizes and two random choices of treatment order. As demonstrated in Supplementary Table 8, different treatment orders by which the joint treatment weights were estimated did not lead to appreciable or directional changes in the estimate of $\bm{\psi}$. However, more efficiency was gained by using the decreasing order of treatment sizes (first conditioning on the treatment used by most patients), as suggested by the narrowest confidence intervals of $\bm{\psi}$.

\section{Discussion}\label{sec:discussion}
Motivated by the need for and inconclusive real-world evidence for the comparative effectiveness of multiple treatments for COVID-19, we have developed a joint marginal structural survival model and novel weighting schemes to address time-varying confounding and censoring in continuous time. There are three main advantages of our proposed method. First, this approach casts the complex time-varying treatment with irregular ``start/stop'' switches into the process of recurrent events where treatment initiation can be considered under the recurrent event framework with discontinuous intervals of eligibility. This innovative formulation enables us to address complex time-varying confounding by modeling the intensity processes of the filtered counting processes for complex time-varying treatments. Second, the proposed method is able to handle a complex longitudinal dataset on its own terms, without discretizing and artificially aligning measurement times, which would lead to less accurate and efficient treatment effect estimates, as demonstrated by our simulations. Third, modern machine learning techniques designed for censored survival data and smoothing techniques of the baseline intensity  can be used easily with our weighting method to further improve the treatment effect estimator under conventional parametric formulations. We have also introduced a simulation algorithm that is compatible with the complex data structures of our proposed modeling framework, and demonstrated the accuracy of the proposed method for estimating causal parameters. Since the proposed method is inspired by large-scale electronic health records data on COVID-19, it is well-suited for sizable longitudinal datasets. Optimal performance can be achieved with a sample size of at least 1000.

We acknowledge several limitations of our study that could merit future investigations. First, we have primarily focused on developing and implementing the continuous-time weighting scheme to address confounding bias with complex observational data. Due to the complexity in the construction of weights, we have only empirically explored convergence rates in simulations. A more in-depth theoretical investigation into the convergence rates of continuous-time weighted JMSSM is warranted in future research, particularly when the weights are estimated by machine learning algorithms. A promising direction to pursue is to explore the possibility of deriving an efficient influence function for our JMSSM \citep{hines2022demystifying}, which has been demonstrated as a vehicle to integrate machine learners for nuisance functions such that the final estimator achieves the standard root-$n$ parametric rate for causal inference \citep{chernozhukov2018double}. These existing results, however, are typically based on a time-fixed treatment, and an extension of those theoretical results to time-varying treatments and continuous-time data would be particularly instrumental and can suggest ways to further improve our proposed estimators. Second, our developments have not addressed the potential challenge arising from competing events. Due to lack of information on specific causes of death in the motivating COVID-19 dataset, we have considered the time to in-hospital death (overall death) as a composite outcome of interest and answered the question on comparative effectiveness on overall mortality. When scientific interest lies in the counterfactual cumulative incidence on patient death due to a specific cause, our JMSSM may not be directly applicable without modifications. One potential approach is to consider a structural model based on the cumulative incidence, mimicking the existing development with a time-fixed treatment \citep{ozenne2020estimation}. Assuming a time-fixed treatment and discrete-time survival data, \citet{young2020causal} provided an inspiring discussion on different versions of causal estimands with competing risks data, and provided recipes for estimation based on g-formula and inverse probability weighting. It would also be useful to investigate the extensions of those results to time-varying treatments and possibly continuous-time survival data. Third, our analysis of COVID-19 dataset has assumed a correctly specified marginal structural proportional hazards model \citep{hernan2001marginal}. Under the proportional hazards assumption, the structural regression parameters can be causally interpreted as they can be derived as contrasts defined on log counterfactual survival functions \citep{martinussen2020subtleties}. In more general settings without proportional hazards, however, the hazard ratio parameters are challenging to interpret \citep{hernan2010hazards}. To partially account for such challenges in interpretation, we have also considered the structural model as a working model and focused on interpreting the resulting counterfactual survival functions under specific treatment regimens, and assessed the robustness of results under an alternative additive marginal structural model. Furthermore, we have also not addressed effect modification by baseline covariates. From a methodological perspective, including baseline effect modifiers is straightforward under our framework. However, in the absence of further content knowledge (especially in the COVID-19 context), identification of baseline effect modifiers in the presence of multiple treatments is challenging and recent studies have only started such explorations in completed randomized trials \citep{blette2023causal}. As treatment effect heterogeneity is out of the scope of our work, we anticipate a further study that aims to specifically investigate effect modification with longotudinal observational data. Finally, we have maintained the conditional exchangeability assumption in our work. This is a standard causal structural assumption in the literature on addressing time-varying confounding.  Although untestable using the observed data, our clinician investigators reflected on the validity of the assumption in the COVID-19 dataset upon the review of time-varying confounders. For future research, expanding the methodology for addressing baseline unmeasured confounding \citep{hogan2014bayesian, hu2022flexible} and developing sensitivity analysis approaches to capture the impact of time-varying unmeasured confounding in continuous time for our model would be a worthwhile and important contribution.

\section{Supplementary Material}
Supplementary sections, tables and figures referenced in Section~\ref{sec:CT-IPW}, \ref{sec:sim}, \ref{sec:analysis} can be found in the Web Supplementary Material. 

\section*{Data availability}
$\R$ codes to implement the proposed methods and replicate our simulation studies are provided in the GitHub page of the first author \url{https://github.com/liangyuanhu/JMSSM-CT}. Access to the COVID-19 data used in the case study needs to be requested and approved by the Icahn School of Medicine at Mount Sinai.

\section*{Funding}
This work was supported in part by award ME-2021C2-23685 from the Patient-Centered Outcomes Research Institute (PCORI) and by grant 1R01HL159077-01A1 from the National Institute of Health (NIH). The statements presented in this article are solely the responsibility of the authors and do not necessarily represent the official views of PCORI\textsuperscript{\textregistered}, its Board of Governors or Methodology Committee or the NIH.

\section*{Conflict of Interest}
None declared.

\bibliographystyle{jasa3}
\bibliography{covid}

\section*{Figure legends}
\end{document}


\begin{center}
\vspace*{-2.5cm}

{\Large Web-based Supplementary Materials for ``Estimating the causal effects of multiple intermittent treatments with application to COVID-19" by Hu et al.}

\medskip
\blinding{
Liangyuan Hu \quad \quad Jiayi Ji \quad \quad Himanshu Joshi \quad \quad Erick Scott \quad \quad Fan Li 

}{}

\end{center}

\date{}

\section{Additional technical details}
\subsection{The kernel function estimator}
In Section 3.4, four approaches are described to estimate the stabilized time-varying inverse probability of treatment weights.  The Nelson-Aalen estimator of the baseline intensity is smoothed by means of kernel functions in approach (ii) and approach (iv). For exposition brevity, consider a simple multiplicative intensity model for treatment initiation $$\rho(t) = \rho_0 (t) \text{IR}\lp \bar{L}(t),\theta \rp Y(t),$$
where $\rho_0 (t)$ is the baseline intensity rate function,  $\text{IR}\lp \bar{L}(t),\theta \rp > 0$ is the (time-dependent) intensity ratio function parameterized by $\theta$ (note that $\text{IR}\lp \bar{L}(t),\theta \rp $ can be nonparametrically estimated via machine learning techniques such as the random survival forests\citep{yao2022ensemble}), and $Y(t)$ is the at risk indicator. The cumulative baseline intensity is $\Rho_0(t) = \int_0^t \rho_0(s)ds$. The kernel function estimator for $\rho_0 (t)$ is derived by smoothing the increments of the Nelson-Aalen estimator of $\hat{\Rho}_0$ as 
\[
\hat{\rho}_0(t)= b^{-1} \int_{\mathscr{T}} K\lp \dfrac{t-s}{b} \rp d\hat{\Rho}_0(s),
\]
where $K$ is the kernel function, which is  a bounded function on $[-1,1]$ and has integral 1, and the bandwidth $b$ is a positive parameter governing the amount of smoothness. Commonly used kernel functions include the standard normal density function $K(x) = \phi(x)$ and the Epanechnikov kernel function $K(x) = 0.75 (1-x^2), \; |x| \leq 1$. \cite{andersen1993statistical} shows that the kernel function estimator $\hat{\lambda}_0(t)$ is consistent and asymptotically normal provided that  there exists a sequence of positive constants $\{a_n\}$, increasing to infinity as $n \rightarrow \infty$, and that the bandwidth tends to zero  more slowly than $a_n^{-2}$ as $n \rightarrow \infty$. 

Placed in the framework of recurrent event for a treatment $A_w$ that can have multiple ``start/stop''  switches as described in Section 3.1,  the intensity of the $q$th treatment initiation $\rho^{A_w}_{q}(t)$ is smoothed by means of kernel function estimator of the baseline intensity function $\rho_{0q}^{A_w}(t)$.    

In our simulations (Section 5) and COVID-19 case study (Section 6), we used both the standard normal density $\phi(x)$ and Epanechnikov kernel functions and they yielded similar results.  We presented results from the normal density kernel function. Following \cite{andersen1993statistical}, we chose the optimal bandwidth $b$ as the value that minimizes the mean integrated squred error (MISE) 
\[
\text{MISE}(\hat{\rho}_0) = E\int_0^{t^o} \lsq \hat{\rho}_0 (t) - \rho(t) \rsq^2 dt.
\]

\subsection{Random survival forests accommodating time-varying covariates}
Approach (iii) uses a recent random survival forests model \citep{yao2022ensemble} that accommodates time-varying covariates to reduce the parametric assumptions about the form of the intensity ratio function required by the usual proportional intensity regression model. \cite{yao2022ensemble} proposed a forest method that estimates the survival function by three steps. In step (1),  we reformat the data in the counting process structure, that is, for individual $i$, the time-varying covariate $L^{(i)}(t)$ will be represented as  $L^{(i)}(t) = l_j^{(i)}$, $t \in \lsq t_j^{(i)}, t_{j+1}^{(i)}\rp $, $j = 0, \ldots, J^{(i)} -1 $. Then split the individual $i$ observation into $J^{(i)}$ pseudo subject observations: $\lp t_j^{(i)}, t_{j+1}^{(i)}, \delta_j^{(i)}, l_j^{(i)}\rp$ with left-truncated right-censored (LTRC) times $t_j^{(i)}, t_{j+1}^{(i)}$ and event indicator $\delta_j^{(i)}$ for the time interval $\lsq t_j^{(i)}, t_{j+1}^{(i)}\rp$. Pool the counting process styled records from $N$ subjects to create a list of pseudo-subjects,
\begin{equation} \label{eq: pseduo}
    \lbc t_l^{'}, t_{l+1}^{'}, \delta_l^{'}, l_l^{'} \rbc_{l=1}^n, \;  n = \sum_i^N J^{(i)}. 
\end{equation}
The set of pseudo-subjects is treated as if they were independent. Step (2) is to apply the forest algorithms on the reformatted dataset given in~\eqref{eq: pseduo}, to fit a model. In step (3), given a particular stream of covariate values 
at the corresponding time values, a survival function estimate is constructed based on the outputs of the forest algorithms. 

We now briefly describe the forest algorithms. \cite{yao2022ensemble} extended the relative risk forests, which combines the relative risk trees \citep{leblanc1992relative} with random forest methodology \citep{breiman2001random}, for LTRC data by modifying the splitting criteria.  The splitting criterion under the relative risk framework is to maximize the reduction in the one-step deviance between the log-likelihood of the saturated model and the maximized log-likelihood. Let $\mathcal{R}_h$ denote the set of observations that fall into the node $h$. Let  $\Rho_0$ index the baseline cumulative intensity function,  $\varphi_h$ represent the nonnegative relative risk of the node $h$, and $t_l$ and $\delta_l$ be the time and event indicator of the $l$th observation  $\forall l\in \mathcal{R}_h$.   Given the right censored observations $(t_l, \delta_l)$,  the full likelihood deviance residual for node $h$ is defined as
\begin{equation} \label{eq:deviance}
    d_h = \sum_{l \in \mathcal{R}_h} 2 \lbc  \delta_l \log \lp  \dfrac{\delta_l}{\hat{\Rho}_0(t_l) \hat{\varphi}_h} \rp  - \lp \delta_l -\hat{\Rho}_0(t_l) \hat{\varphi}_h \rp\rbc.
\end{equation}
Two steps are needed to modify the splitting rule and obtain the deviance residual appropriate for LTRC data \citep{fu2017survival}. First, compute the estimated cumulative intensity function $\hat{\Rho}_0 (\cdot)$ based on all pseudo-subject observations. Second, replace $\hat{\Rho}_0 (t_l)$ in~\eqref{eq:deviance} with $\hat{\Rho}_0 (t_{l+1}^{'}) - \hat{\Rho}_0 (t_{l}^{'})$, and replace $\delta_l$ in~\eqref{eq:deviance} with $\delta_l^{'}$. We refer to \cite{yao2022ensemble} for more detailed description of the random survival forests model. 

\subsection{Variance estimation}


Since our estimators for $\bm{\psi}$ (marginal structural proportional hazards model parameter) can be considered as solution to the weighted partial score equation, we consider the robust sandwich variance estimator as a convenience device to construct confidence intervals. The robust sandwich variance estimator has been considered, for example, in \citet{xiao2010accuracy}, and \citet{karim2017application,karim2018comparison}, and has been shown to be at most conservative under the discrete-time setting. We use this estimator for inference in conjunction with our continuous-time stabilized inverse probability weights, and formally evaluate its performance in our simulations. We briefly describe the robust sandwich variance estimator below. With the continuous-time weights, we focus on equation (11) in the main text with two treatments:
\begin{equation*}
   \sum_{i=1}^n \int_0^\infty \Omega^{A_1,A_2} \Omega^C(G_i)  \lbc Z(A_{1i},A_{2i},t) - \frac{S^{(1)}(t;\bm{\psi})}{S^{(0)}(t;\bm{\psi})} \rbc dN_i^T(t)=0, 
\end{equation*}
where $\Omega^{A_1,A_2} \Omega^C(G_i)$ is the weight for time-varying treatments $A_1$ and $A_2$ and censoring (in continuous time), $Z(A_{1i},A_{2i},t)_{(3\times 1)} = [A_{1i}(t), A_{2i}(t), A_{1i}(t)A_{2i}(t)]^{\top}$, and
\begin{align*}
S^{(0)}(t;\bm{\psi})=&\sum_{k \in \mathcal{R}_t^T} Y^{\ast  \ast T}_{k}(t) r(A_{k1},A_{k2},t ;\bm{\psi}) \\
S^{(1)}(t;\bm{\psi})=&\sum_{k \in \mathcal{R}_t^T} Z(A_{k1}, A_{k2}, t) Y^{\ast  \ast T}_{k}(t) r(A_{k1},A_{k2},t ;\bm{\psi}).
\end{align*}
In the above definition, $\mathcal{R}_t^T$ refers to the collection of subjects still at risk for the outcome event at time $t$, $Y^{\ast  \ast T}_{k}(t) = \Omega^C(G_i) \Omega^{A_1,A_2}(t_{K_i}) Y^T_k(t)$, where $Y_i^T(t)$ is the at-risk function for the outcome event, and $r(a_1,a_2,t) = \exp \{\psi_1 a_1(t) + \psi_2 (t) a_2(t) + \psi_3 a_1(t) a_2(t)\} $. Now further define
\begin{equation*}
S^{(2)}(t;\bm{\psi})=\sum_{k \in \mathcal{R}_t^T} Z(A_{k1}, A_{k2}, t)^{\otimes 2} Y^{\ast  \ast T}_{k}(t) r(A _{k1},A_{k2},t ;\bm{\psi}),
\end{equation*}
with $a^{\otimes 2}=aa^{\top}$ for any vector $a$. 
Then the robust sandwich variance estimator takes the form $\bm{\Sigma}_0^{-1}\bm{\Sigma_1}\bm{\Sigma}_0^{-1}$ \citep{binder1992fitting}, where
\begin{equation*}
\bm{\Sigma}_0=\sum_{i=1}^n \int_0^\infty \Omega^{A_1,A_2} \Omega^C(G_i)  \lbc \frac{S^{(2)}(t;\bm{\psi})}{S^{(0)}(t;\bm{\psi})} - \frac{S^{(1)}(t;\bm{\psi})^{\otimes 2}}{S^{(0)}(t;\bm{\psi})} \rbc dN_i^T(t),
\end{equation*}
and 
\begin{equation*}
\bm{\Sigma}_1=\sum_{i=1}^n \lsq \int_0^\infty \Omega^{A_1,A_2} \Omega^C(G_i)  \lbc Z(A_{1i},A_{2i},t) - \frac{S^{(1)}(t;\bm{\psi})}{S^{(0)}(t;\bm{\psi})} \rbc dM_i^T(t) \rsq^{\otimes 2},
\end{equation*}
and $M_i^T(t)=N_i^{T}(t)-\int_0^t Y^{\ast  \ast T}_{k}(u)\lambda_0(u) r(A _{k1},A_{k2},u ;\bm{\psi})du$ is the martingale based on the counting process for outcome event. The sandwich variance estimator is obtained when both $\bm{\Sigma}_0$ and $\bm{\Sigma_1}$ are evaluated at the estimated weights and $\hat{\bm{\psi}}$. 

Because the above robust variance estimator considers the weights $\Omega^{A_1,A_2} \Omega^C(G_i)$ as fixed known values \citep{binder1992fitting}, it could result in conservative (but still valid) inference. With time-invariant weights estimated by logistic regression in the cross-sectional treatment setting, the corrected robust sandwich variance estimator has been derived to achieve improved variance estimation for hazard ratio parameters \citep{shu2021variance}. However, an extension to continuous-time weights is not trivial and currently unavailable. Alternatively, resampling method such as the bootstrap method \citep{brumback2004sensitivity,cole2008constructing} could be used to make more robust inference for $\bm{\psi}$ that accounts for the uncertainty of the weights. 

\section{Marginal Structural Cox Model Simulation Algorithm}

Here we provide pseudocodes for the marginal structural cox model data simulation. We use the COVID-19 dataset as the foundation to set the values of the parameters in both the treatment assignment and marginal structural models. In our simulations, to reduce the impact of data sparsity, we set the maximum number of treatment initiations to be 4. To do this, the pseudocodes can be modified by setting treatment exposure status to one at all time points after the fourth treatment initiation throughout to the end of the follow-up period. 
\vspace{0.4cm}

\hrule

\vspace{0.4cm}

 \textbf{GET} 
 
 $\quad M \leftarrow 100$ (maximum follow-up);
 $\quad \lambda_{0}\leftarrow 0.005;$
 $\quad n \leftarrow 1000;$

$\quad \bm{\beta} \leftarrow(\beta_{0},\beta_{1},\beta_{2},\beta_{3})\leftarrow[\log(3/7),-0.5,-\log(1/2),\log(3/2)]$ (parameter vector for generating time-varying confounding variables $L_2$);

$\quad \bm{\zeta} \leftarrow(\zeta_{0},\zeta_{1},\zeta_{2},\zeta_{3},\zeta_{4})\leftarrow[\log(2/7),-\log(1/2),-0.5, \log(3/2), \log(2/3)]$ (parameter vector for generating time-varying confounding variables $L_1$);

$\quad \bm{\gamma}\leftarrow(\gamma_{0},\gamma_{1},\gamma_{2},\gamma_{3},\gamma_{4},\gamma_{5},\gamma_{6},\gamma_{7},\gamma_{8},\gamma_{9},\gamma_{10},\gamma_{11})$

\;\;\;\;$\quad$$\leftarrow[\log(2/7),1/2,-1/2,-\log(3/5),0.8,0.5,0.8,-0.5,1/2,1.2,-0.6,-0.3] $ (parameter vector for generating $A_1$);

$\quad$$ \bm{\eta}\leftarrow(\eta_{0},\eta_{1},\eta_{2},\eta_{3},\eta_{4},\eta_{5},\eta_{6},\eta_{7},\eta_{8},\eta_{9},\eta_{10},\eta_{11})$ 

\;\;\;\;$\quad\leftarrow[\log(3/7),1/3,-1/3,-\log(2/5),0.9,0.6,0.8,-0.5,1/3,0.9,-0.6,-0.4]$ (parameter vector for generating $A_2$);

$\quad  \psi_{1}\leftarrow-0.5$ (true log-hazard value representing the 
effect of treatment $A_1$ )

$\quad \psi_{2}\leftarrow-0.3$ (true log-hazard
value representing the effect of treatment $A_2$)

\textbf{COMPUTE}

 \textbf{for} ID $ \gets 1$ to $n$ (for each individual) \textbf{do}

 $\quad$INIT: $L_1(-1)\leftarrow 0;  L_2(-1)\leftarrow 0; A_1(-1)\leftarrow 0;  A_2(-1)\leftarrow 0; Y(0)\leftarrow 0; H(m)\leftarrow 0;  $
 
 $\quad \quad \quad \quad  n_{A_1} \leftarrow 0; n_{A_2} \leftarrow 0; \tau_{A_1} \leftarrow 0; \tau_{A_2} \leftarrow 0$
 
 $\quad T^{\bar{0}}\sim \text{Exponential}(\lambda_{0})$
 
 
$\quad$ \textbf{for} $m \gets 0$ to M  \textbf{ do}

$\quad \quad L_1(m) \leftarrow E\lp L_1(m) = l_1(m) \cond L_1(m-1), A_1(m-1), A_2(m-1), Y(m)=0; \bm{\zeta} \rp $

$\hspace{1.8cm} \quad \leftarrow \zeta_0 + \zeta_1(1/\log T^{\bar{0}}) + \zeta_2 A_1(m-1) + \zeta_3 L_1(m-1) + \zeta_4 A_2(m-1) $

$\quad \quad \text{logit}(p_{L_2})  \leftarrow \text{logit} P\lp L_2(m)=1\cond L_2(m-1),A_1(m-1),A_2(m-1),Y(m)=0;\bm{\beta}\rp$

$\hspace{2.3cm}\quad\leftarrow\beta_{0}+\beta_{1}A_1(m-1)+\beta_{2}L_2(m-1)+\beta_3 A_2(m-1) $

$\quad \quad L_2(m) \sim \text{Bernoulli}(p_{L_2})$

$\quad \quad \text{logit}(p_{A_1})  = \text{logit} P ( A_1(m)=1\cond L_1(m),L_1(m-1),L_2(m),L_2(m-1),A_1(m-1),$ 

$\hspace{3.3cm}   A_2(m-1),n_{A_1}, Y(m)=0;\bm{\gamma} )$

$\hspace{2.8cm}  =\gamma_{0}+\gamma_{1}A_1(m-1)+\gamma_2(L_2(m-1))^{2}+\gamma_{3}(L_1(m-1))^{2} +\gamma_{4}(A_1(m-1)L_1(m))$

$\hspace{2.5cm} \quad +\gamma_{6}(L_1(m)L_2(m))+\gamma_{7}(A_1(m-1)L_2(m))+\gamma_{8}A_2(m-1) $

$\hspace{3cm} +\gamma_{9}(A_2(m-1)L_1(m))+\gamma_{10}(A_2(m-1)L_2(m))+\gamma_{11}n_{A_1}$

$\quad \quad \textbf{if }  A_1(m-1) = 0  \text{ or } m-1=\tau_{A_1}$  \textbf{  then }

$\quad \quad \quad  A_1(m)  \sim \text{Bernoulli}(p_{A_1})$

$\quad \quad \quad  \textbf{if }   A_1(m) = 1 $ \textbf{ then }

$\quad \quad \quad \quad \delta_{A_1} \sim \text{zero-truncated Poisson} (10)$ (treatment duration after initiation)

$\quad \quad \quad \quad \tau_{A_1} \leftarrow m + \delta_{A_1}$

$\quad \quad \quad \quad A_1(m+1): A_1(\max{(\tau_{A_1}, M)}) \leftarrow 1$ 

$\quad \quad \quad \quad n_{A_1} \leftarrow n_{A_1} + 1$


$\quad \quad \quad  \textbf{end if}$

$\quad \quad  \textbf{end if}$

$\quad \quad  \text{logit}(p_{A_2})  = \text{logit} P ( A_2(m)=1\cond L_1(m),L_1(m-1),L_2(m),L_2(m-1),A_1(m),$ 

$\hspace{3.2cm}   A_2(m-1),n_{A_2},Y(m)=0;\bm{\eta} )$

$\hspace{2.8cm}  =\eta_{0}+\eta_{1}A_1(m-1)+\eta_2(L_2(m-1))^{2}+\eta_{3}(L_1(m-1))^{2} +\eta_{4}(A_1(m)L_1(m))$

$\hspace{2.8cm}  +\eta_{6}(L_1(m)L_2(m))+\eta_{7}(A_1(m)L_2(m))+\eta_{8}A_2(m-1) $

$\hspace{2.8cm} +\eta_{9}(A_2(m-1)L_1(m))+\eta_{10}(A_2(m-1)L_2(m))+\eta_{11}n_{A_2}$

$\quad \quad \textbf{if }  A_2(m-1) = 0  \text{ or } m-1=\tau_{A_2}$ \textbf{ then }

$\quad \quad \quad A_2(m)  \sim \text{Bernoulli}(p_{A_2})$

$\quad \quad \quad \textbf{if }  A_2(m) = 1$  \textbf{then}

$\quad \quad \quad \quad \delta_{A_2} \sim \text{zero-truncated Poisson} (9)$ (treatment duration after initiation)

$\quad \quad \quad \quad \tau_{A_2} \leftarrow m + \delta_{A_2}$

$\quad \quad \quad \quad A_2(m+1): A_2(\max{(\tau_{A_2}, M)}) \leftarrow 1$ 

$\quad \quad \quad \quad n_{A_2} \leftarrow n_{A_2} + 1$

$\quad \quad \quad  \textbf{end if}$

$\quad \quad  \textbf{end if}$

$\quad \quad H_{m} \leftarrow H_m + \exp \lbc \psi_1 A_1(m) + \psi_2 A_2(m) \rbc $

$\quad \quad  \textbf{if } T^{\bar{0}} \geq H_m$ 

  $\quad \quad \quad  Y_{m+1} \leftarrow 0$

$\quad \quad  \textbf{else} $

$\quad \quad \quad  Y_{m+1} \leftarrow 1 $

$\quad \quad \quad  T \leftarrow m + (T^{\bar{0}} - H_m) \times \exp \lbc -\psi_1 A_1(m) - \psi_2 A_2(m)\rbc $

$\quad \quad  \textbf{end if}$

$\quad  \textbf{end for m}$

$\textbf{end for \text{ID}}$

\vspace{0.4cm}

\hrule

\vspace{0.4cm}

\section{Additional  details of recurrent events formulation}\label{sec:recurrent}
To formalize the treatment initiation process, we first consider a univariate treatment process $N^{A_w}$. We assume that the jumps of $A_w(t)$, i.e., $dA_w(t)$, is observed on certain subintervals of $[0,t^o]$ only. Specifically for individual $i$, we observe the stochastic process $A_{w,i}(t)$ on a set of intervals 
$$\mathcal{E}_{w,i} = \bigcup_{j=1}^{J_i} (V_{w,ij}, U_{w,ij}],$$
where $0 \leq V_{w,i1} \leq U_{w,i1} \leq \ldots \leq V_{w,iJ_i} \leq U_{w,iJ_i} \leq t_{w,iK_i}$. This representation implies the following results. First, an individual can have at most $J_i \geq 1 $ initiations of treatment $w$:  if $U_{w,iJ_i} = t_{w,iK_i}$, then individual $i$ has $J_i-1$ treatment initiations; and if $U_{w,iJ_i} < t_{w,iK_i}$, then individual $i$ has $J_i$ treatment initiations. A special case where $J_i=1$ and $U_{w,iJ_i} = t_{w,iK_i}$ corresponds to the situation where individual $i$ is continuously eligible for treatment initiation and has not been treated with $w$ during the follow-up. Second, once treatment is initiated, $A_{w,i}(t)$ is no longer stochastic until person $i$ discontinues the treatment. This also suggests that the $j$th treatment initiation is observed at $U_{w,ij}$. This implication pertains to the ``off'' treatment period (referred to as the lack-of-color period in Figure 1), during which a patient becomes eligible to receive the treatment.  Third, we have 
$$A_{w,i}(t) =1,~~~\forall~t~~~\in (U_{w,ij}, V_{w,i(j+1)}], j=1, \ldots, J_i-1.$$ In words, treatment status is equal to one deterministically on the discontinuous intervals of ineligibility (i.e., \emph{on} treatment period). This implication involves the 
 ``on'' treatment period (represented by the colored segment in Figure 1). Once the treatment commences at a specific time point, the patient continues with the treatment for a predetermined duration, throughout which the treatment status remains deterministically at one.

\section{A justification for consistency using the Radon-Nikodym derivative}
This section offers a heuristic rationale for the treatment weights employed in fitting our structural proportional hazards model. We employ arguments analogous to those presented by \cite{hu2018modeling} and \cite{johnson2005semiparametric}, who utilized a Radon-Nikodym (R-N) derivative developed by \cite{murphy2001marginal} to construct and estimate weights that lead to consistent estimates of structural parameters in the context of censored survival outcomes and continuous outcomes, respectively. For brevity, we demonstrate the use of R-N assuming no censoring. For the scenario involving right censoring, the reasoning follows a similar approach.

Under randomization of treatment, the unbiased partial likelihood score equations can be written as $\sum_i^n D_i(\bm{\psi}) = 0$ \citep{fleming2013counting, hu2018modeling}, where 
$$D_i(\bm{\psi}) = \int_0^\infty  \lbc Z(A_{1i},A_{2i},t) - \bar{Z}(t;\bm{\psi}) \rbc dN_i^T(t)=0,$$ $Z(A_{1i},A_{2i},t)_{(3\times 1)} = [A_{1i}(t), A_{2i}(t), A_{1i}(t)A_{2i}(t)]^{\top}$, and
$$\bar{Z} = \dfrac{\sum_{k \in \mathcal{R}_t^T} Z(A_{k1}, A_{k2}, t) Y_{k}(t) r(A_{k1},A_{k2},t ;\bm{\psi})}{\sum_{k \in \mathcal{R}_t^T}Y_{k}(t) r(A_{k1},A_{k2},t ;\bm{\psi}) }. $$

Now we consider the scenario in which treatment is nonrandomly allocated. Let $\mathbbm{P}_R(\cdot)$ denote the data distribution under randomized treatment, and let $\mathbbm{P}_O(\cdot)$ denote the same under non-random allocation of treatment. The observable data, with respect to treatment and outcome, for each individual, under either randomized or non-randomized allocation of treatment, is $\{\bar{A}_1(T^*),\bar{A}_2(T^*), T^*, \Delta^T\}$. Following \cite{murphy2001marginal}, under the 
 conditional exchangeability assumption (A2) and some regularity conditions, including the Positivity assumption (A3), the distribution of $\{\bar{A}_1(T^*), \bar{A}_2(T^*), T^*, \Delta^T\}$ under $\mathbbm{P}_R(\cdot)$ is absolutely continuous with respect to the distribution of $\{\bar{A}_1(T^*),\bar{A}_2(T^*), T^*, \Delta^T\}$ under $\mathbbm{P}_O(\cdot)$, and a version of the R-N derivative is 
\begin{eqnarray} \label{eq:R-N}
    E_o \lbc \dfrac{f^{A_1, A_2}\lp T^*\rp}{ f^{A_1,A_2} \lp T^* \cond \bar{L}\lp T^* \rp \rp } \cond T^*, \Delta^T, \bar{L}(T^*) \rbc, 
\end{eqnarray} 
where $f^{A_1,A_2}(\cdot)$ is the joint probability density according to which treatments $A_1$ and $A_2$ are randomly allocated,  and $f^{A_1,A_2}(\cdot \cond \cdot)$ is the conditional joint probability density.

An estimating equation that is a function of observed
data and is unbiased under the distribution of $\mathbbm{P}_R(\cdot)$ can be
re-weighted by the R–N derivative to obtain an unbiased estimating
equation using the same observed data, but now under
the distribution $\mathbbm{P}_O(\cdot)$ \citep{murphy2001marginal}. We apply the R-N derivative in~\eqref{eq:R-N} to construct an unbiased estimating equation using the observed data under the  distribution $\mathbbm{P}_O(\cdot)$, by noticing that the partial likelihood estimating equation $\sum_i^n D_i(\bm{\psi})$ can be represented as three averages. We first rewrite $D_i(\bm{\psi})$ as 
$$D_i(\bm{\psi}) = \int_0^\infty  \lbc Z(A_{1i},A_{2i},t) - \dfrac{n^{-1}\sum_{k=1}^n Z(A_{k1}, A_{k2}, t) Y_{k}(t) r(A_{k1},A_{k2},t ;\bm{\psi})}{n^{-1} \sum_{k=1}^n Y_{k}(t) r(A_{k1},A_{k2},t ;\bm{\psi}) } \rbc dN_i^T(t)=0. $$

Under randomized assignment, 
\begin{eqnarray}\label{eq:exp-1}
    n^{-1}\sum_{k=1}^n Z(A_{k1}, A_{k2}, t) Y_{k}(t) r(A_{k1},A_{k2},t ;\bm{\psi}) &\overset{P}{\to}& E_R \lbc Z(A_{k1}, A_{k2}, t) Y_{k}(t) r(A_{k1},A_{k2},t )\rbc \nonumber \\
    &=& M_1(t; \bm{\psi}),
\end{eqnarray}

\begin{eqnarray}\label{eq:exp-2}
    n^{-1}\sum_{k=1}^n  Y_{k}(t) r(A_{k1},A_{k2},t ;\bm{\psi}) &\overset{P}{\to}& E_R \lbc Y_{k}(t) r(A_{k1},A_{k2},t )\rbc \nonumber \\
    &=& M_0(t; \bm{\psi}), 
\end{eqnarray}

and 

\begin{eqnarray}\label{eq:exp-3} 
    n^{-1}\sum_{i=1}^n \int_0^\infty  \lbc Z(A_{1i},A_{2i},t) - \dfrac{M_1(t; \bm{\psi})}{M_1(t; \bm{\psi})} \rbc dN_i^T(t) \nonumber \\
    \overset{P}{\to} E \lbc \int_0^\infty  \lbc Z(A_{1i},A_{2i},t) - \dfrac{M_1(t; \bm{\psi})}{M_1(t; \bm{\psi})} \rbc dN_i^T(t)  \rbc. 
\end{eqnarray}

Each of the three limits in \eqref{eq:exp-1}, \eqref{eq:exp-2} and \eqref{eq:exp-3} is an expected value, taken over the distribution of treatment,  applied to observable data under randomization. We can apply the  R-N derivative in~\eqref{eq:R-N} to \eqref{eq:exp-1}, \eqref{eq:exp-2} and \eqref{eq:exp-3}, and obtain the weights for both the risk set and the score contribution. The weighted weighted partial score equation is 
\begin{equation*} 
     \sum_{i=1}^n \int_0^\infty \Omega^{A_1,A_2}(t_{K_i})  \lbc Z(A_{1i},A_{2i},t) - \bar{Z}^{\ast }(t;\bm{\psi}) \rbc dN_i^T(t)=0, 
\end{equation*}
where the general form of the weight is $\Omega^{A_1,A_2}(t_{K_i}) = \dfrac{f^{A_1, A_2}\lp T^*\rp}{ f^{A_1,A_2} \lp T^* \cond \bar{L}\lp T^* \rp \rp } $. 

The continuous-time is a generalization of the discrete-time weight. For example, in the discrete-time setting with (two) nonrecurrent treatments, the stabilized inverse probability weights (we suppress subscript $i$ for brevity) are given in the prior literature \citep{howe2012estimating,hernan2001marginal}: 
\begin{align}\label{eq:dis-weight}
\begin{split}
    \Omega^{A_1,A_2}(t) = & \lbc \prod_{\{k: t_k \leq t\}} \dfrac{P \lp A_1(t_k)=a_1(t_k) \cond \bar{A}_1({t_{k-1}}), \bar{A}_2({t_{k-1}}) \rp }{P \lp A_1(t_k) = a_1(t_k) \cond \bar{A}_1({t_{k-1}}), \bar{A}_2({t_{k-1}}), \bar{L}(t_{k-1}), T \geq t, C \geq t  \rp} \rbc \times  \\
    & \lbc \prod_{\{k: t_k \leq t\}} \dfrac{P \lp A_2(t_k) = a_2(t_k) \cond \bar{A}_1({t_{k}}), \bar{A}_2({t_{k-1}}) \rp }{P \lp A_2(t_k)=a_2(t_k) \cond \bar{A}_1({t_{k}}), \bar{A}_2({t_{k-1}}), \bar{L}(t_{k-1}), T \geq t, C \geq t  \rp}\rbc,
    \end{split}
\end{align}
where $t_k$'s are a set of ordered discrete time points common to all individuals satisfying $0=t_0 < t_1 <t_2<\ldots \leq t$. In Section 3.2, we generalize these weights to the continuous-time setting, and further take into account the recurrent nature of the treatments.

\section{Precise formulation of individual weights}
When the number of time intervals in $[0,t]$ increases and $ds$ approaches zero, the finite product over the number of time intervals of the individual partial likelihood will approach a product integral \citep{aalen2008survival}. 
Therefore, as we discuss in the main manuscript, each product term in the numerator and denominator of \eqref{eq:dis-weight} can be generalized to the continuous-time setting as
\begin{eqnarray} \label{eq:prod_int}
    &&\Prodi_0^t { \lbc D^{A_w}(s) \lambda^{A_w}(s \cond \bullet )ds  \rbc ^{ dA_w(s) }  \lbc D^{A_w}(s) \lp 1-\lambda^{A_w}(s \cond \bullet )ds \rp  \rbc ^{ 1-dA_w(s) }}   \nonumber\\
    &=& \lsq \Prodi_0^t {\lbc D^{A_w}(s) \lambda^{A_w}(s \cond \bullet )  \rbc ^{ \Delta A_w(s) }} \rsq \exp \lbc -\int_0^t  D^{A_w}(s) \lambda^{A_w}(s \cond \bullet )ds \rbc,
\end{eqnarray}
where $\Delta A_w(t) = A_w(t)-A_w(t^-)$.  For individual $i$, both factors in~\eqref{eq:prod_int} need to be evaluated with respect to the individual's filtered counting process $\{N_{iq}^{A_w}(t): 0 \leq t \leq t_{K_i}, q =1, \ldots, Q_{w,i}\}$, 
where $Q_{w,i}$ is the number of initiations of treatment $w$ for individual $i$.

As described in Supplementary Section~\ref{sec:recurrent}, the number of treatment initiations for individual $i$,  $Q_{w,i}$ can take three values: (i) $Q_{w,i}=0$, (ii) $Q_{w,i} = J_i-1$ or (iii) $Q_{w,i} = J_i$. Corresponding to the three cases, the quantity in~\eqref{eq:prod_int} can be rewritten in explicit forms as
 $$
\text{Quantity}~\eqref{eq:prod_int} =\begin{cases}
			S^{A_w}\lp t_{K_i} \cond \bullet\rp & \text{if $Q_{w,i}=0$}\\
            f^{A_w}\lp U_{i,J_i-1} \cond \bullet\rp \lbc S^{A_w} \lp  V_{iJ_i} \cond \bullet \rp - S^{A_w}\lp t_{iK_i} \cond \bullet \rp \rbc  & \text{if $Q_{w,i}=J_i-1$}\\
        f^{A_w}\lp U_{iJ_i}  \cond \bullet\rp  & \text{if $Q_{w,i}=J_i$},   
		 \end{cases}
$$
where $S^{A_w}$ and $f^{A_w}$ are the survival and density function of the filtered counting process for treatment $A_w$. In alignment with conventions established in prior literature \citep{hernan2001marginal,howe2012estimating}, the exposure weight for joint treatments $A_1$ and $A_2$ assumes a treatment order. By positing that treatment $A_1$ is administered infinitesimally earlier than treatment $A_2$, the intensity of initiating treatment $A_2$ at time $t$ can be dependent on the status of treatment $A_1$ at time $t$, $A_1(t)$, and the status of treatment $A_2$ at time $t^-$,  $A_2(t^-)$. Furthermore, the intensity of initiating treatment $A_1$ at time $t$ can rely on $A_1(t^-)$ and $A_2(t^-)$. This assumption of ordered treatment administration is plausible in clinical contexts.

For exposition brevity, we define
\begin{align*}
&\bar{\mathcal{O}}_1(t) = \{\bar{A}_1(t^-), \bar{A}_2(t^-),  \bar{L} (t^-),T \geq t, C \geq t \}\\
&\bar{\mathcal{O}}_2(t) = \{\bar{A}_1(t), \bar{A}_2(t^-),  \bar{L} (t^-),T \geq t, C \geq t \}\\
&\bar{\mathcal{O}}^{A_1}(t) = \{\bar{A}_1(t^-), \bar{A}_2(t^-), T\geq t, C\geq t\}\\
&\bar{\mathcal{O}}^{A_2}(t) = \{\bar{A}_1(t), \bar{A}_2(t^-), T\geq t, C\geq t\}
\end{align*}
Putting this all together, the individual continuous-time stabilized inverse probability weight that corrects for time-varying confounding is given by $\Omega^{A_1,A_2}(t) = \Omega^{A_1}(t)\Omega^{A_2}(t)$ with $\Omega^{A_w}$ being: 
\vspace{-3pt}
\begin{align} \label{eq:ct-ipw} 
\begin{split} 
  & \Omega^{A_w}(t_{K_i}) \\
=&  \begin{cases}
     \dfrac{S^{A_w}\lp t_{K_i} \cond \bar{\mathcal{O}}^{A_w}(t_{K_i})\rp}{S^{A_w}\lp t_{K_i} \cond \bar{\mathcal{O}}_w(t_{K_i}) \rp} & \text{if $Q_{w,i}=0$}\\[3mm]
    \dfrac {f^{A_w}\lp U_{i,J_i-1} \cond \bar{\mathcal{O}}^{A_w}(U_{i,J_i-1}) \rp \lbc S^{A_w} \lp  V_{iJ_i} \cond \bar{\mathcal{O}}^{A_w}(V_{iJ_i}) \rp - S^{A_w}\lp t_{iK_i} \cond \bar{\mathcal{O}}^{A_w}(t_{K_i}) \rp \rbc} {f^{A_w}\lp U_{i,J_i-1} \cond \bar{\mathcal{O}}_w(U_{i,J_i-1}) \rp \lbc S^{A_w} \lp  V_{iJ_i} \cond \bar{\mathcal{O}}_w(V_{iJ_i}) \rp - S^{A_w}\lp t_{iK_i} \cond \bar{\mathcal{O}}_w(t_{K_i}) \rp \rbc }  & \text{if $Q_{w,i}=J_i-1$}\\[3mm]
    \dfrac{f^{A_w} \lp U_{iJ_i} \cond \bar{\mathcal{O}}^{A_w}(U_{iJ_i}) \rp }{ f^{A_w} \lp U_{iJ_i} \cond \bar{\mathcal{O}}_w(U_{iJ_i}) \rp} & \text{if $Q_{w,i}=J_i$} 
    \end{cases}
    \end{split}
\end{align}

\section{Supplementary figures and tables}
\subsection{Additional simulation results}


\begin{figure}[h!b!p!]
    \centering
    \includegraphics[width=0.9\textwidth]{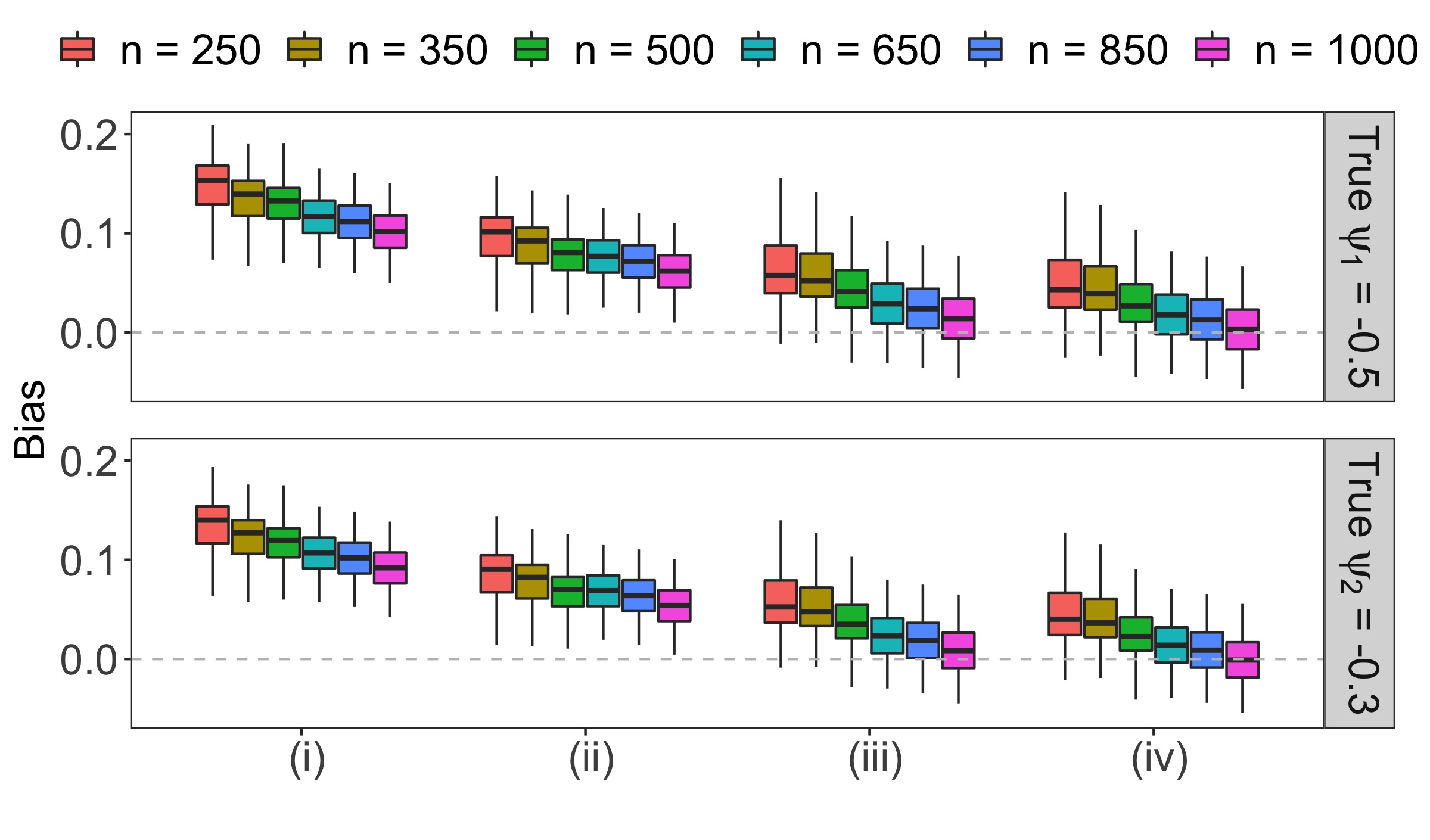}
    \caption{Biases in estimates of $\psi_1$ and $\psi_2$ for sample sizes $n=$ 250, 350, 500, 650, 850, and 1000 across 250 data replications, using four weight estimation approaches detailed in Section 3.4. Approach (i) uses main-effects Cox regression model and Nelson-Aalen estimator for baseline intensity. Approach (ii) uses kernel function smoothing of the Nelson-Aalen estimator in approach (i). Approach (iii) uses a survival forests model that accommodates time-varying covariates and Nelson-Aalen estimator for baseline intensity. Approach (iv) uses kernel function smoothing of the Nelson-Aalen estimator in approach (iii). }
\end{figure}

\begin{figure}[h!b!p!]
    \centering 
    \includegraphics[width=1\textwidth]{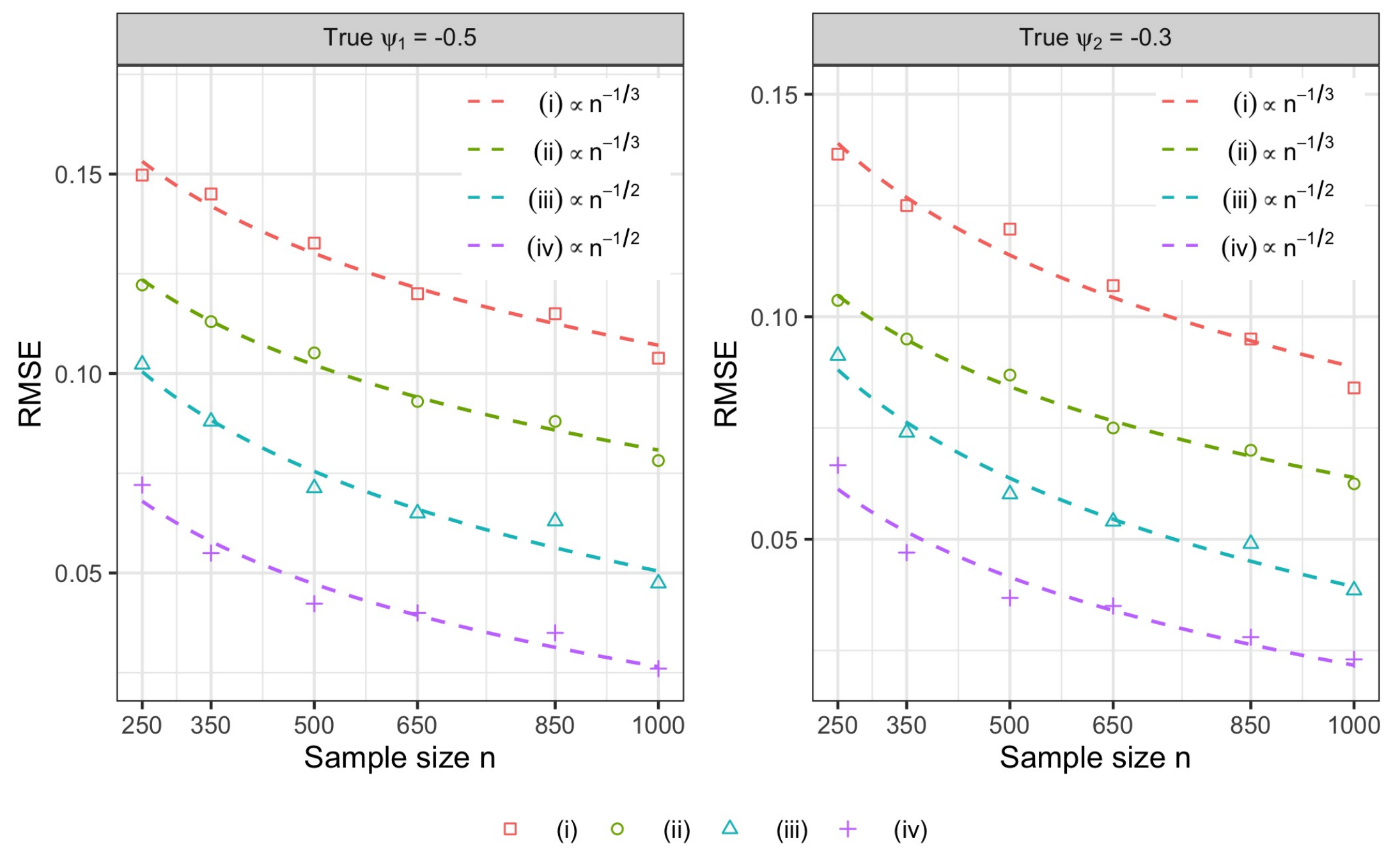}
    \caption{Approximate convergence rates of root-mean-square errors for estimating structural parameters $\psi_1$ and $\psi_2$ utilizing the four weighting estimators detailed in Section 3.4. }
\end{figure}

\begin{figure}[h!b!p!]
    \centering
    \includegraphics[width=1\textwidth]{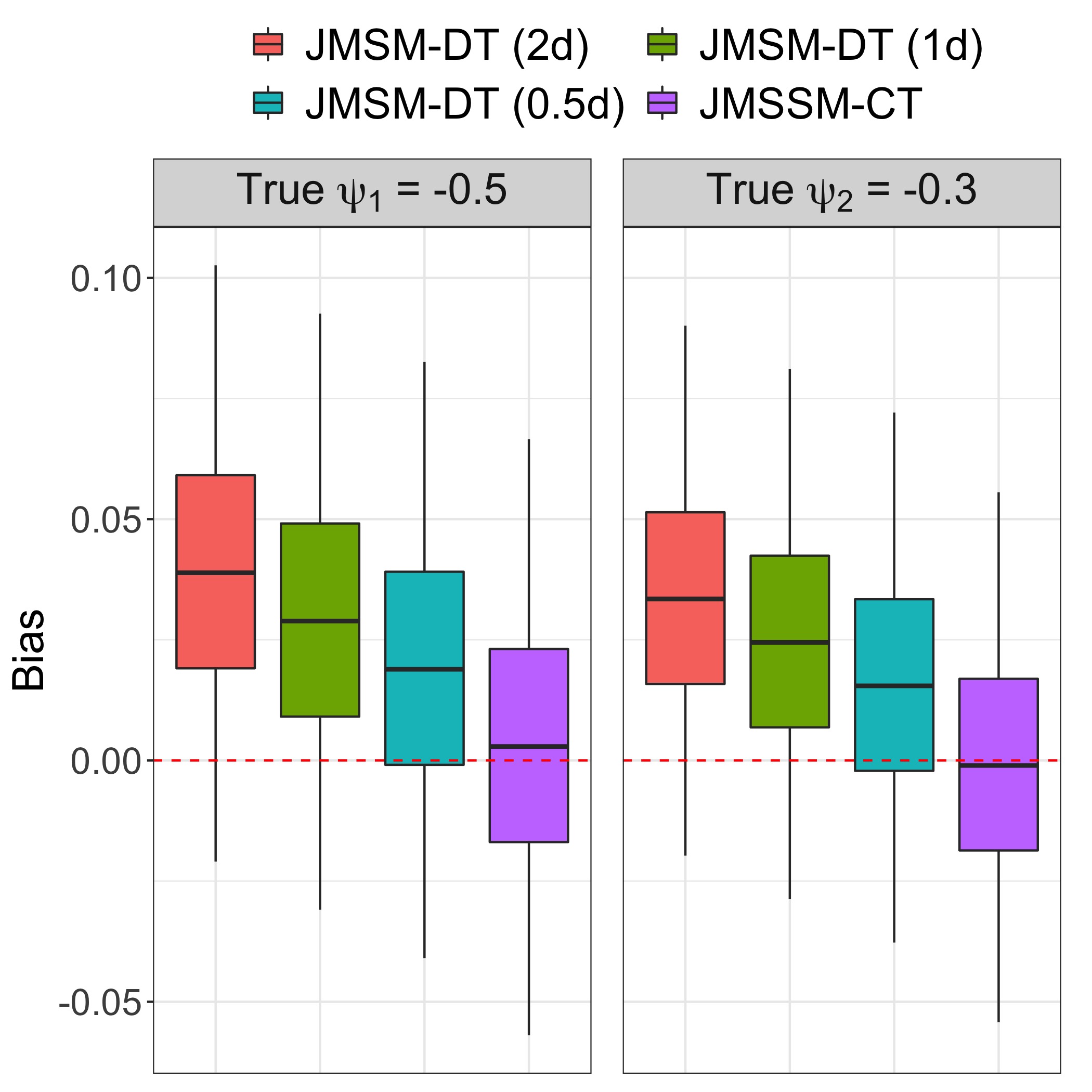}
    \caption{The distributions of biases across 250 simulation replications of the  ragged longitudinal data with unaligned time points, in estimating the parameters $\psi_1$ and $\psi_2$ using the proposed JMSSM-CT method and the comparison method JMSM-DT. When implementing JMSM-DT, the follow-up time was respectively discretized into time intervals of length 0.5, 1 and 2 days, as the method requires aligned measurement time points. }
    \label{fig:bias_JSMMSCTvsJMSMDT}
\end{figure}

\begin{figure}[h!b!p!]
    \centering
    \includegraphics[width=0.8\textwidth]{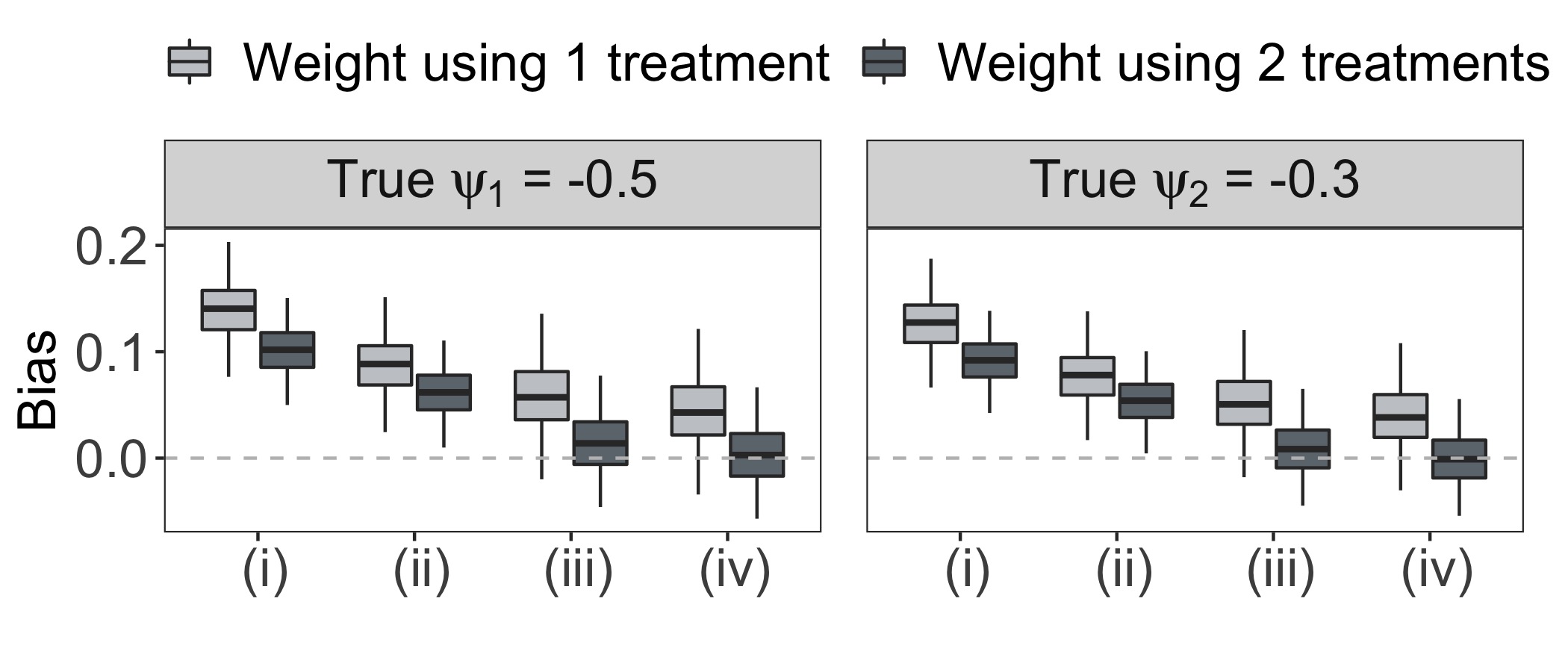}
    \caption{Biases in the estimate of $\bm{\psi}$ using two types of weighting estimators:  one focusing on estimating a single treatment effect and one utilizing joint treatment weights.}
\end{figure}

\begin{figure}[h!b!p!]
    \centering
    \includegraphics[width=1\textwidth]{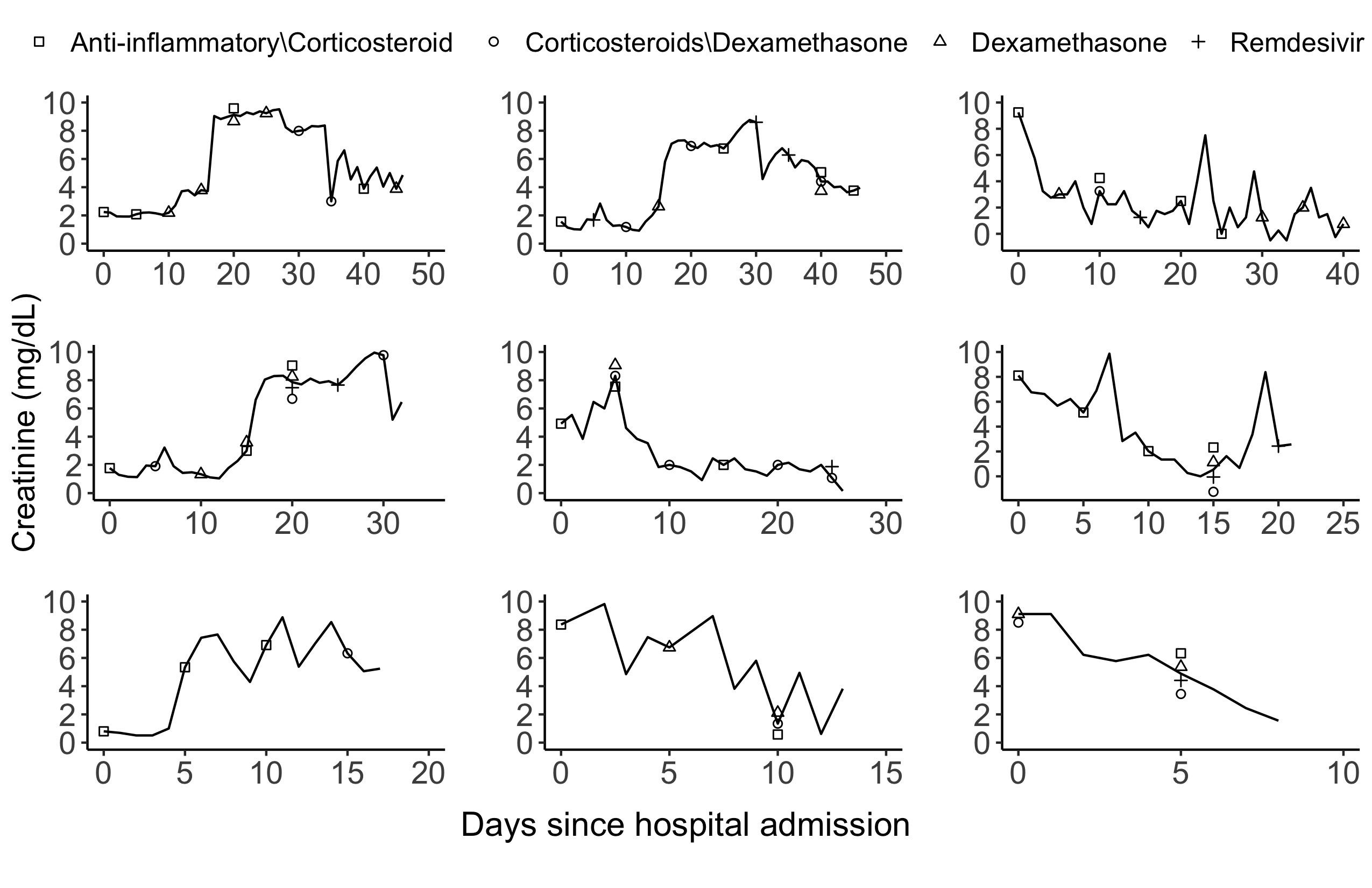}
    \caption{Trajectories of serum creatinine levels over the course of hospital stay for 9 randomly chosen patients. Symbols represent the types of treatment classes received by a patient at a given time.}
\end{figure}

\begin{figure}[h!b!p!]
    \centering
    \includegraphics[width=1\textwidth]{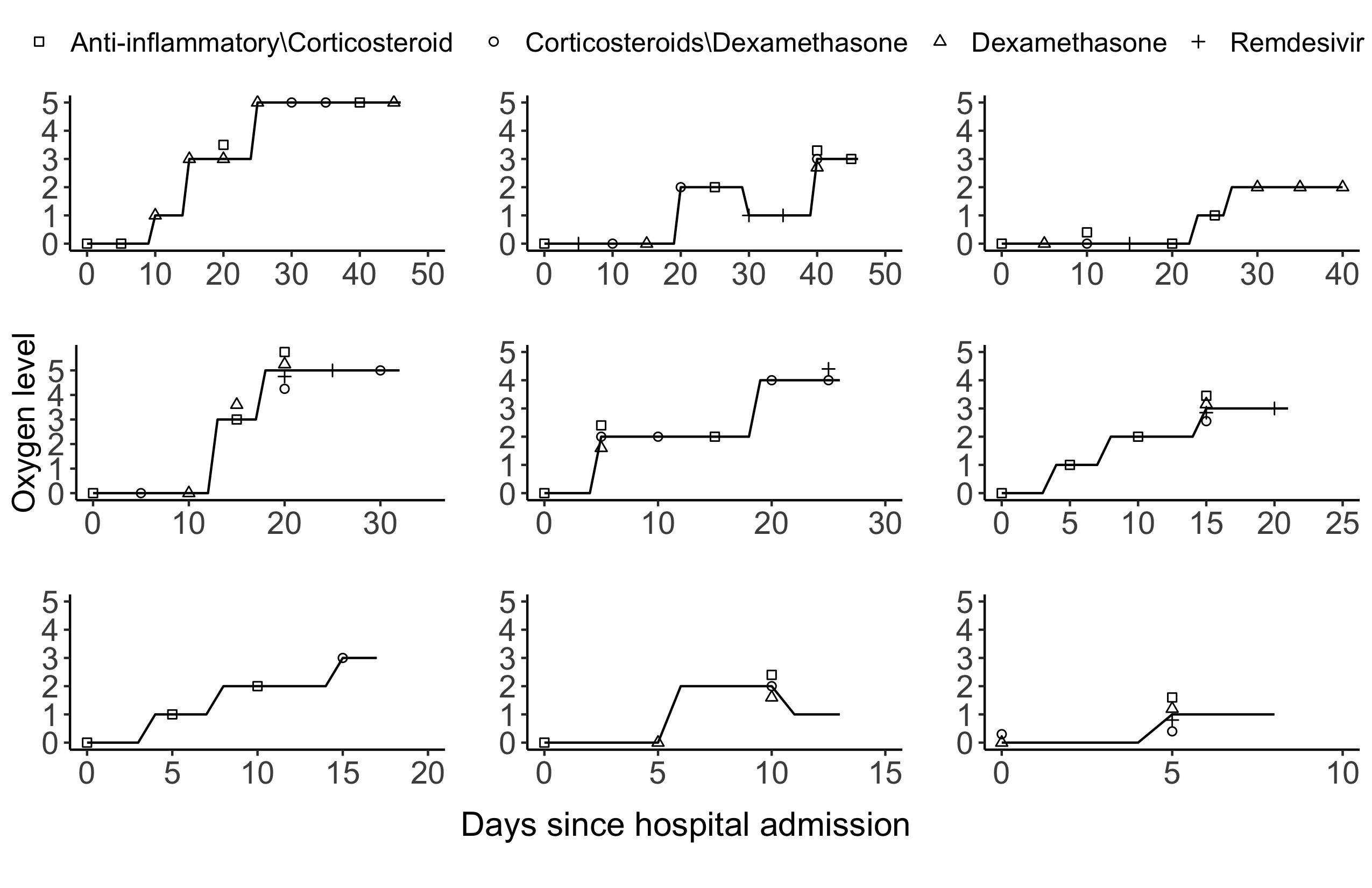}
    \caption{Trajectories of patient oxygen levels over the course of hospital stay for 9 randomly chosen patients. Symbols represent the types of treatment classes received by a patient at a given time.}
\end{figure}

\begin{table}[H]
\centering
\caption{Mean absolute bias (MAB), root mean square error (RMSE) and coverage probability (CP) for the estimates of $\bm{\psi}$ across 250 data replications with unaligned follow-up time points, using four weighting estimators (i)-(iv) described in Section 3.4. }
\begin{tabular}{clcccccccccccc}
\toprule
\multirow{2}{*}{weighting estimators}&  \multicolumn{3}{c}{$\psi_1$} && \multicolumn{3}{c}{$\psi_2$}\\
\cline{2-4} \cline{6-8}
& MAB & RMSE & CP && MAB & RMSE & CP  \\
\midrule
(i) & .102 & .104   & .048 && .092 &.094   &  .060 \\ 
(ii) & .063 &.067  &  .104 && .055& .059 &   .112 \\ 
(iii) & .023 & .029 &    .936 && .020 & .025 &   .940 \\ 
(iv) & .016 &.022  &  .952 && .015& .019 &   .952 \\ 
\bottomrule
\end{tabular}
\label{tab:4_weight_estimator}
\end{table}

\begin{table}[H]
\centering
\caption{The distribution of the estimated individual time-varying weights from one random replication of the ``ragged'' longitudinal data with unaligned time points, for the proposed JMSSM-CT versus JMSM-DT. To estimate the weights, the random forests was used for JMSM-DT and four approaches (i)-(iv) (Section 3.4) were used for JMSSM-CT.  }
\begin{tabular}{ccccccccc}
\toprule
\multirow{2}{*}{Methods}&\multirow{2}{*}{weighting estimators}&\multicolumn{5}{c}{Distribution of estimated weights}\\ \cline{3-7}
& & Minimum & First quartile & Mean & Third quartile & Maximum   \\
\midrule
\multirow{4}{*}{JMSSM-CT}&(i) & 0.23 & 0.89   & 1.05 & 1.23 & 5.34    \\ 
&(ii) & 0.40 & 0.98   & 1.01 & 1.07 & 4.28    \\ 
&(iii) & 0.52 & 0.88   & 1.00 & 1.09 & 2.99   \\ 
&(iv) & 0.68 & 0.95   & 1.00 & 1.05 & 2.36   \\ 
JMSM-DT &Random forests & 0.43 & 0.90    &  1.03 & 1.11  & 3.65   \\ 
\bottomrule
\end{tabular}
\label{tab:weight_dist}
\end{table}

\clearpage 

\subsection{Treatment classes for COVID-19}
In Table~\ref{tab:Covid_trt}, we provide detailed definitions of the four treatment classes for COVID-19 whose comparative effectiveness on in-hospital death was investigated in Section 6. 
\begin{table}[H] 
\centering
  \caption{Definitions of four treatment classes for COVID-19. iv:intravenous; po: by mouth.} 
 \def\arraystretch{1.25} 
\begin{tabular}{>{\raggedright\arraybackslash}p{4cm} p{12cm}} 
\toprule 
Treatment classes   & Medication  (route) \\ 
\midrule
Dexamethasone & Dexamethasone (iv), Dexamethasone (po)\\
Remdesivir & Remdesivir (iv)\\
Corticosteroids other than dexamethasone & Hydrocortisone (po), Hydrocortisone (iv), Methylprednisolone (po), Methylprednisolone (iv), Prednisolone (iv), Prednisone (po), Prednisone (iv)\\
Anti-inflammatory medications other than corticosteroids & Alpha-1-Proteinase Inhibitor (iv), Anakinra (iv), Azathioprine (po), Belatacept (iv), Eculizumab (iv), Envarsus (iv), Sarilumab (iv), Gengraf (po), Gengraf (iv), Hydrocortisone (po), Hydrocortisone (iv), Ibrutinib (po), Immune Globulin (iv), Infliximab (iv), Methylprednisolone (po), Methylprednisolone (iv), Montelukast (po), Prednisolone (iv), Prednisone (po), Prednisone (iv), Ruxolitinib (po), Tocilizumab (iv), Apremilast (po), Celecoxib (po), Dasatinib (po), Everolimus (iv), Gemtuzumab (iv), Ibuprofen (iv), Ibuprofen (po), Ifosfamide (iv), Leflunomide (po), Mesalamine (iv), Methotrexate (iv), Methylphenidate (iv), Mycophenolate (iv), Naproxen (po), Rituximab (iv), Sulfasalazine (po), Tacrolimus (iv), Pacritinib (po), Risankizumab (iv), Daratumumab (iv), Talquetamab (iv)\\
 \bottomrule
\end{tabular}
 \label{tab:Covid_trt}
\end{table}
\clearpage 

\subsection{Patient oxygen levels}
We describe how patient oxygen levels are categorized based on the use of ventilator in Table~\ref{tab:def_pol}. 

\begin{table}[H]
 \caption{Definitions of patient oxygen levels based on the use of ventilator} 
 \def\arraystretch{1.25} 
\begin{tabular}{p{3.5cm} p{12.5cm}}
\toprule
Patient oxygen level   & Ventilator status \\
 \midrule
0 & Room air\\
1 & Cannula\\
2 & Mask, Blow-by, Face tent, Oxyhood, Non-rebreather, RAM cannula\\
3 & Continuous positive airway pressure machine, High flow nasal cannula, Hudson prongs\\
4 & Bilevel positive airway pressure machine, Tracheostomy mask\\
5 & Tracheotomy, Transtracheal oxygen therapy, Ventilator, Endotracheal tube, T-shaped tubing connected to an endotracheal tube, Nasal synchronized intermittent mandatory ventilation\\
 \bottomrule
\end{tabular}\label{tab:def_pol}
\end{table}


\subsection{Additional COVID data analysis results}
  
 The inclusion criteria for the COVID-19 dataset are as follows: individuals aged 18 years or older, confirmed COVID-19 diagnosis via a positive reverse-transcriptase PCR test conducted on a nasopharyngeal swab, and hospital admission at one of the five hospitals within the Mount Sinai Health System between February 25, 2020, and February 26, 2021. No exclusion criteria were applied.

\begin{table*}[t]
\centering
\footnotesize
 \def\arraystretch{0.6} 
\caption{Baseline characteristics of patients from the COVID-19 data. Summary statistics are represented as mean (standard deviation [SD]) for continuous variables and No. (\%) for discrete variables.}
\begin{tabular}{lc}
\toprule 
 Characteristics & $N = 11286$  \\
  \hline
  Age (years), mean (SD) & 64.59 (18.16) \\  Gender N ($\%$) &\\
  \;\;  Male &    6137 (54.4)       \\ 
  \;\;  Female &  5149 (45.6)   \\  
  Race, N ($\%$) &\\
 \;\;White & 3241 (28.7)    \\ 
 \;\;Black &  2829 (25.1)   \\ 
 \;\;   Asian  &    647 (5.7)    \\ 
  \;\;   Others  &    4569 (40.5)    \\ 
 Ethnicity N ($\%$) &\\
 \;\;  Hispanic &  2972 (26.3)     \\ 
  \;\; non-Hispanic &  8314 (73.7)   \\ 
Asthma N ($\%$) &\\
  \;\;  Yes &    1082 (9.6)       \\ 
  \;\;  No &  10204 (90.4)  \\ 
  COPD N ($\%$) &\\
  \;\;  Yes &      734 (6.5)      \\ 
  \;\;  No &  10552  (93.5) \\ 
    Hypertension N ($\%$) &\\
  \;\;  Yes &       5593 (49.6)      \\ 
  \;\;  No &   5693 (50.4) \\ 
   Cancer N ($\%$) &\\
  \;\;  Yes &        1161 (10.3)      \\ 
  \;\;  No &   10125  (89.7) \\ 
  Coronary artery disease N ($\%$) &\\
  \;\;  Yes &         2227 (19.7)       \\ 
  \;\;  No &    9059 (80.3) \\ 
  Diabetes N ($\%$) &\\
  \;\;  Yes &          2809 (24.9)  \\ 
  \;\;  No &   8477 (74.1) \\ 
  Smoking status N ($\%$) &\\
   \;\;  Current &   598 (5.3)  \\
  \;\;  Former &   2597 (23.0)    \\ 
  \;\;  Never &    6350 (56.3)  \\ 
  \;\;  Unknown &   1741 (15.4)    \\ 
  Hospital site N ($\%$) &\\
   \;\;     Mount Sinai Brooklyn &    1671 (14.8)   \\
  \;\;     Mount Sinai Petrie &     771 (6.8)     \\ 
  \;\;   Mount Sinai Queens &     1565 (13.9)   \\ 
  \;\;     Mount Sinai St. Luke's &  1828 (16.2)     \\ 
  \;\;   Mount Sinai West &     1304 (11.6)   \\ 
  \;\;   Mount Sinai Main Hospital &    4147 (36.7)     \\ 
 \bottomrule
\end{tabular}
\footnotesize \\\smallskip
Abbreviations:  SD = standard deviation; 
\end{table*}

\begin{table*}[t]
\centering
\caption{Comparing the proposed JMSSM-CT with discrete-time based method JMSM-DT in estimating the joint and interactive effects  $\hat{\bm{\psi}}$ (log hazard ratio) of COVID-19 treatments and associated 95\% confidence intervals (CI), using the COVID-19 dataset drawn from  the  Epic  electronic  medical  records  system  if  the  Mount  Sinai  Medical  Center. The composite outcome of in-hospital death or admission to ICU was used. Confidence intervals were estimated via the robust sandwich variance estimators. ``$\times$" denotes treatment interaction. The weighting estimator (iv) was used for JMSSM-CT. For JMSM-DT,  the follow up time was discretized in the space of 1, 3, and 5 days. The smallest unit of time in the COVID-19 data is 1 day.} 
\vspace{0.1in}
\def\arraystretch{1.1}
\scalebox{0.9}{
\hspace*{-3em}
\begin{tabular}{lcccc}
\toprule
\multirow{2}{*}{Treatment orders}&\multicolumn{4}{c}{ $\hat{\bm{\psi}}$ (95\% Confidence Interval)} \\
\cline{2-5}
 & JMSSM-CT & JMSM-DT (1d) & JMSM-DT (3d) & JMSM-DT (5d) 
  \\
\midrule
Dexamethasone & $-.20 (-.35, -.06)$ & $-.22 (-.39, -.05)$ &   $-.25(-.45, -.05)$ & $-.27(-.55, .01)$\\ 
Remdesivir  & $-.53 (-.75, -.31)$ & $-.50 (-.76,-.25)$ &   $-.48(-.76, -.20)$& $-.42(-.74, -.10)$ \\
Corticosteroids other than & & &&\\\smallskip 
 dexamethasone & $-.08 (-.29, .19)$ & $-.10 (-.36, .21)$ &  $-.12 (-.43, .22)$ & $-.14 (-.48, .24)$ \\
Anti-inflammatory medications & & &&\\\smallskip 
other than corticosteroids & $-.05 (-.56, .47)$ & $-.07 (-.60, .48)$&   $-.09(-.66, .49)$& $-.13 (-.75,.51)$\\
Remdesivir $\times$ Corticosteroids & & &&\\\smallskip 
other than dexamethasone & $-.74 (-.95, -.52)$ & $-.78 (-1.01, -.55)$ &   $-.81(-1.07, -.56)$ & $-.83 (-1.14, -.52)$   \\
\bottomrule 
\end{tabular}
\label{tab:discrete-time-covid}
}
\end{table*}

\begin{table*}[t]
\centering
\caption{Causal inferences about the treatment effects of COVID-19 medications on the bases of 14-day counterfactual survival probability and 14-day restricted mean survival time (RMST), using proposed JMSSM-CT with weighting estimator (iv) with additive hazard outcome model. The composite outcome of in-hospital death or admission to ICU was used. The 95\% Confidence intervals were estimated using nonparametric bootstrap with 100 repetitions.} 
\vspace{0.1in}
\def\arraystretch{1.1}
\scalebox{0.9}{
\hspace*{-3em}
\begin{tabular}{lcccc}
\toprule
Treatment & 14-day counterfactual survival probability & 14-day counterfactual RMST\\ 
\midrule
Dexamethasone &  0.860 (0.835, 0.882) & 13.11 (12.84, 13.49)\\ 
Remdesivir  &  0.892 (0.870, 0.902) & 13.12 (12.85, 13.48)\\
Corticosteroids other than & & &&\\\smallskip 
 dexamethasone & 0.832 (0.812, 0.849) & 12.63 (12.38, 12.97) \\
Anti-inflammatory medications & & &&\\\smallskip 
other than corticosteroids & 0.814 (0.783, 0.833) & 12.74 (12.32, 13.07)\\
Remdesivir + Corticosteroids & & &&\\\smallskip 
other than dexamethasone &  0.912 (0.881, 0.939) & 13.21 (12.98, 13.52)\\
\bottomrule 
\end{tabular}
\label{}
}
\end{table*}

\begin{table*}[!htbp]
\centering
\caption{The joint and interactive effect estimates $\hat{\bm{\psi}}$ (log hazard ratio) of COVID-19 treatments and associated 95\% confidence intervals (CI), using the COVID-19 dataset drawn from  the  Epic  electronic  medical  records  system  of  the  Mount  Sinai  Medical  Center. The composite outcome of in-hospital death or admission to ICU was used. The weighting estimator (iv) (Section 3.3) was used. The joint treatment weights were estimated using four different choices of treatment order. Confidence intervals were estimated via the robust sandwich variance estimators. ``$\times$" denotes treatment interaction. $A_1 =$ dexamethasone. $A_2 = $ corticosteroids other than dexamethasone. $A_3 = $ remdesivir. $A_4 = $ anti-inflammatory medications other than corticosteroids. $\rightarrow$ denotes treatment order. For example,  $A_1$ $\rightarrow$ $A_2$ $\rightarrow$ $A_3$ $\rightarrow$ $A_4$ indicates that the joint treatment weights $\Omega^{A_1, A_2, A_3, A_4}$ are estimated by $\Omega^{A_1, A_2, A_3, A_4} = \Omega^{A_1} \times \Omega^{A_2 \cond A_1}  \times \Omega^{A_3 \cond A_1,A_2}  \times \Omega^{A_4 \cond A_1,A_2,A_3}$. }
\vspace{0.1in}
\def\arraystretch{1.1}
\setlength{\tabcolsep}{1pt}
\scalebox{0.9}{
\hspace*{-3em}
\begin{tabular}{lcccc}
\toprule
\multirow{2}{*}{Treatment orders}&\multicolumn{4}{c}{ $\hat{\bm{\psi}}$ (95\% Confidence Interval)} \\
\cline{2-5}
 & $A_1$ $\rightarrow$ $A_2$ $\rightarrow$ $A_3$ $\rightarrow$ $A_4$ & $A_4$ $\rightarrow$ $A_3$ $\rightarrow$ $A_2$ $\rightarrow$ $A_1$ & $A_1$ $\rightarrow$ $A_3$ $\rightarrow$ $A_4$ $\rightarrow$ $A_2$& $A_4$ $\rightarrow$ $A_2$ $\rightarrow$ $A_1$ $\rightarrow$ $A_3$ 
  \\
\midrule
Dexamethasone & $-.20 (-.35, -.06)$ & $-.22(-.43, -.01)$ &   $-.23(-.38, -.08)$ & $-.21 (-.41, -.02) $\\ 
Remdesivir  & $-.53 (-.75, -.31)$ & $-.48(-.74, -.22)   $&   $-.56(-.79, -.33)  $& $-.50 (-.74,-.26)$\\
Corticosteroids other than & & &&\\\smallskip                                        
 dexamethasone & $-.08 (-.29, .19)$ &$-.02(-.33, .29)  $ &  $-.12(-.37, .13)  $ & $-.04 (-.34, .26)$ \\
Anti-inflammatory medications & & &&\\\smallskip 
other than corticosteroids & $-.05 (-.56, .47)$ &$-.01(-.59, .57)$  &   $-.08(-.63, .47)$&  $-.03 (-.58, .51)$\\
Remdesivir $\times$ Corticosteroids & & &&\\\smallskip 
other than dexamethasone & $-.74 (-.95, -.52)$ &$-.65(-.92, -.38)$  &   $-.80(-1.05, -.55)$ &  $-.68 (-.94, -.44)$ \\ 
\bottomrule
\end{tabular}}
\label{tab:order_covid}
\end{table*}

\clearpage
\bibliographystyle{jasa3} 
\bibliography{covid}